# *De novo* visual proteomics in single cells through pattern mining

Min Xu[1], Elitza I Tocheva[4], Yi-Wei Chang[2], Grant J Jensen[2,3], and Frank Alber[1,*]

*1 Molecular and Computational Biology, Department of Biological Sciences, University of Southern California, Los Angeles, California, USA.*

*2 Division of Biology and Biological Engineering, California Institute of Technology, Pasadena, California, USA.*

*3 Howard Hughes Medical Institute, Pasadena, California, USA.*

*4 Department of Stomatology, Université de Montréal, Montreal, Quebec, CANADA.*

** Corresponding author*

**Cryo-electron tomography enables 3D visualization of cells in a near native state at molecular resolution. The produced cellular tomograms contain detailed information about all macromolecular complexes, their structures, their abundances and their specific spatial locations in the cell. However, extracting this information is very challenging and current methods usually rely on templates of known structure. Here, we formulate a template-free visual proteomics analysis as a *de novo* pattern mining problem and propose a new framework called "*Multi Pattern Pursuit*" for supporting proteome-scale *de novo* discovery of macromolecular complexes in cellular tomograms *without* using templates of known structures. Our tests on simulated and experimental tomograms show that our method is a promising tool for template-free visual proteomics analysis.**

Nearly every major process in a cell is orchestrated by the interplay of macromolecular assemblies, which often coordinate their actions as functional modules in biochemical pathways. In order to proceed efficiently, this interplay between different macromolecular machines often requires a distinctly nonrandom spatial organization in the cell. Therefore, when modeling complex biological functions, it is crucial to know the structures, abundances and locations of the entire set of large macromolecular complexes. Currently, proteomics studies have explored the component lists of proteins often extracted from a lysed cell population, but little is known about how proteins and their complexes are spatially arranged in a crowded cell at its native state, limiting the plausibility of modeling complex biological functions.

With cryo-electron tomography (ECT), it is now possible to generate 3D reconstructions of cells in hydrated, close to native states at molecular resolutions. New imaging technologies and advances in automation are allowing labs to quickly obtain large numbers of cryo-electron tomograms. It is therefore now possible to undertake a "visual proteomics"[1,2] analysis of large macromolecular complexes in individual cells. However, the detection and structural analysis of unknown macromolecular complexes in tomograms remains very challenging due to a number of factors. First, macromolecular complexes can vary significantly in shape, size, and cellular abundance. Second, identifying individual complexes is significantly more difficult in cellular tomograms than in isolated samples, due to high crowding levels[3]. Third, individual tomograms often have low signal-to-noise ratio (SNR) and low contrast, in particular cellular tomograms, for which the sample is relatively thick (>300nm). In addition, the tomogram image is modulated by the Contrast Transfer Function (CTF) effect. Finally, the limited range of tilt angles leads to partial sampling of images and missing structural components in the Fourier space, resulting in anisotropic resolution and distortions that depend on the orientation of the object relative to the tilt axis (i.e., the missing wedge effect). Therefore, unlike large organelles, which can be detected by visual inspection, the systematic structural classification and recovery of all accessible macromolecular complexes in cellular tomograms is difficult, and can only be ventured with the aid of highly efficient automatic analysis methods.

The pioneering work to quantitatively analyze the spatial organizations of macromolecular complexes in cellular tomograms used "template matching"[e.g. [1,2,4–6]]. This approach uses a given complex's known structure from other high-resolution *in vitro* methods (e.g. X-ray crystallography, NMR, cryo-EM single particle reconstruction) to simulate a ECT reconstruction, the template, which is then used to search for matches in the experimental tomogram. Naturally, this approach is limited to detecting complexes whose 3D structures are already known, which represent only a small fraction of all the complexes in the cell. Assessing the reliability of detected matches is also challenging[7] because the template structure can misfit its targets, either due to conformational changes or additional bound components to the structure *in vivo*, or because the template structure is from a different organism and exhibits a different conformation[8].

To overcome some of these limitations, a few template-free subtomogram averaging[9] and classification [e.g. [10,11]] approaches have been developed recently. By using iterative clustering and averaging these methods classify subtomograms into groups of similar structures. Due to the computationally intensive nature of 3D image processing (especially the subtask of subtomogram alignment), current methods are tailored to datasets usually containing a relatively small number of structural classes, for example multiple conformational or compositional states of a single macromolecular complex [e.g. [10,12–14]]. Thus, if we wish to apply them on a proteome-wide scale, existing template-free approaches have several drawbacks. First, the computational cost is proportional to the number of subtomograms multiplied by the number of classes. In cellular visual proteomics, the set of subtomograms will contain a large number of different complexes. Moreover, to allow for sufficient copy numbers to obtain a high SNR in each class, it is necessary to process a very large number (tens to hundreds of thousands) of subtomograms. Therefore, these applications are computationally extremely demanding[15]. To our knowledge, no template-free subtomogram classification method exists that is specifically optimized for and can be applied to large-scale applications in a visual proteomics setting.

Here, we address this problem through *Pattern Mining*, which searches for high-quality *structural patterns* occurring frequently in a cellular tomogram. Herein, a structural pattern is defined as a set of aligned subtomograms likely to represent a single complex and its density average. We propose a new framework called *Multi Pattern Pursuit* (MPP) (Figure 1a), which is specifically designed for supporting large-scale pattern mining in visual proteomics to detect *de novo* macromolecular complexes of variable shapes and sizes in cellular tomograms, *without* using templates of known structures. MPP produces the shape and abundance of the patterns as well as its position and orientation in the tomogram.

There are substantial differences between MPP and existing template-free classification methods. Our software is specifically designed to handle (i) large numbers of subtomograms extracted from cellular tomograms (tens of thousands of subtomograms); (ii) sets of subtomograms that may include a large number of different structural classes, with widely varying shapes, sizes, and abundances; and (iii) subtomograms extracted from a crowded environment[3], which may include fragments of neighboring complexes as well as the target complex.

MPP is an iterative constrained optimization process that maximizes the total quality of a set of structural patterns that are distinct with respect to their average density and disjoint with respect to their subtomogram membership. That is, MPP strongly discourages its patterns from sharing subtomograms. MPP generates structural patterns from the collection of subtomograms and stores them into a growing pattern library. At each new MPP iteration a new selection of distinct candidate patterns is made from this library, based on the pattern quality. Specifically, we search for the combination of patterns leading to the best overall quality score, while having a maximal coverage of the subtomograms without substantial overlap of subtomogram membership between the selected patterns. The corresponding density averages of the selected candidate patterns then serve as references for updating the rigid transformations of all subtomograms, which in turn leads to newly generated patterns in the following MPP iteration step, which will be added to the growing pattern library. To achieve this goal, MPP relies on a very efficient subtomogram alignment[11] algorithm based on constrained correlation[11,16] and fast rotational matching[17], and an efficient, robust, and flexible parallel architecture[18] that supports high-throughput processing. In addition, MPP contains procedures to optimize patterns by removing likely misassigned subtomograms and also procedures to remove redundant patterns from the library.

Next, we describe the main steps of the MPP method in more detail. Each MPP iteration consists of the following steps (Figure 1a, Supplementary Note 1.1.1):

**(Step 1) Pattern Generation** (Supplementary Note 1.1.3). Each MPP iteration starts by generating patterns consisting of subtomograms likely to contain similar objects. At every iteration, the complete set of subtomograms is processed using their currently assigned rigid transformations (the very first iteration uses randomly assigned transformations). Note that in each iteration, many independent methods can be used to generate alternative sets of patterns, which all will be added to the pattern library. In other words, the MPP framework is an ensemble method, which uses multiple algorithms to obtain better predictive performance. Here, we use three different methods to generate patterns:

**1) Clustering**. We apply k-means clustering of the subtomograms in a reduced dimension space. To reduce the influence of noise and reduce computational cost we propose a method for dimension reduction that includes an imputation-based strategy to account for the missing-wedge effect (Supplementary Note 1.1.3.1) and the large structural heterogeneity of complexes. At early iterations a fixed cluster number $K$ is used that is chosen to be large enough to increase the chance of detecting patterns of relatively low abundance, while being small enough so that there are sufficient numbers of subtomograms in each cluster to generate a meaningful first pattern density average. After an initial set of iterations an automatically determined value for $K$ is used.

**2) Sequential expansion**. Patterns are also generated by a sequential expansion method (Supplementary Note 1.1.3.2), which optimizes the quality of patterns already generated in the previous iterations: for a given pattern, the subtomograms are ranked based on their alignment scores to the pattern average (obtained in MPP step 5 of the previous iteration). Then a score cutoff is searched that maximizes the quality for a newly formed pattern containing only the subset of subtomograms with alignment scores higher than the cutoff. This procedure allows the exclusion of likely misclassified subtomograms. (This method is applied at later iterations of the MPP optimization process).

**3) (Optional) GA-based single pattern pursuit.** We also perform a combinatorial optimization to increase the SFSC quality score of patterns already generated in previous iterations. Selecting the optimal subset of subtomograms that maximizes the SFSC quality score in a pattern is a combinatorial problem. We developed a Genetic Algorithm (GA) to perform such an optimization (Supplementary Note 1.1.3.3). Due to the computational complexity we apply this method only at the final iteration for a selected set of patterns (Figure 1b).

After pattern generation, the subtomograms in all the generated patterns are averaged to generate the pattern densities.

**(Steps 2 and 3) Determine the quality of the patterns** (Supplementary Note 1.1.2) **and expand pattern library.** We quantify the quality of the density average of each pattern as a "Sum of the Fourier Shell Correlation" (SFSC) score estimated by the Spatial Signal to Noise Ratio (SSNR), which takes into account missing wedge effects and can be efficiently and unambiguously computed. All newly generated patterns and their quality score are then added to the pattern library.

**(Step 4) Selecting disjoint set of candidate patterns of high quality from pattern library** (Supplementary Note 1.1.4).

A new selection of distinct candidate patterns is made from the pattern library, based on their combined pattern quality. Specifically, we search for the combination of patterns leading to the best combined quality score, while having a maximal coverage of subtomograms and without substantial overlap in subtomogram membership between the selected patterns. The selected candidate patterns serve as references for subtomogram alignments in the next step. The new selection of distinct patterns is made on the basis of (i) their SFSC score and (ii) the number of subtomograms appearing in more than one of the selected patterns. The selected pattern subset thus maximizes pattern quality while favoring patterns containing uniquely assigned subtomograms. The selection algorithm is iterative and greedy: it adds patterns from the library one by one until no eligible candidate pattern remain in the library. The first pattern taken is the one with the highest quality score. Then the pattern with the next highest quality score is selected *only if* fewer than 1% of its subtomograms are also present in other candidate patterns of the current selection. Due to the flexibility of the MPP framework in future it is easily possible to implement other selection methods.

**(Step 5) Alignment of selected candidate patterns into common reference frames** (Supplementary Note 1.1.5). The density averages of the selected candidate patterns are aligned into common reference frames, which reduces potential biases resulting from their translational and orientational differences in the subsequent dimension reduction step.

**(Step 6) Alignment of all subtomograms against all the candidate patterns.** Next, we calculate the best alignments between each subtomogram and each of the density averages of all selected candidate patterns. For each subtomogram the transformation and alignment scores against each of the pattern density averages are stored.

**(Step 7) Redundant pattern detection and removal** (Supplementary Note 1.1.6). It is possible that two patterns, formed by disjoint sets of subtomograms, represent the same complex. The structural redundancy of two patterns is quantified by a statistical test based on the pairwise alignment scores of the subtomograms against the density averages of each selected patterns. If two candidate patterns are structurally similar, the one with lower quality is considered redundant and removed from both the current candidate pattern selection and the pattern library.

**(Step 8) Storing rigid transformation for each subtomogram**. Finally, for each subtomogram the rigid transformation of the best scoring alignment among all the alignments to all the non-redundant candidate pattern averages is stored. These updated transformations are used as input information for the new pattern generation step in the next MPP iteration.

(**Optional Step 8b for crowded samples**) **Automatic masking of target complexes**. In a cellular tomogram, a subtomogram is often extracted from a crowded environment. Besides the target complex, a subtomogram may contain neighbor structures and regions of background noise that can substantially bias the processing. As an optional component, we introduce a method for automatic target complex segmentation. The pattern density average leading to the highest subtomogram alignment score (from Step 8) is chosen as a seed to generate an adaptive mask on the subtomograms so that MPP can exclude everything outside the target complex. Such adaptive masking is based on a level set based segmentation (Supplementary Note 1.1.7).

The whole process (steps 1-8) is repeated until a new iteration does not generate any new, non-redundant candidate patterns and has therefore converged to a final set of patterns. The outputs are the candidate patterns of the final iteration, the subtomograms assigned to each pattern and their rigid transformations.

**Pre-processing steps (optional)** (Figure 1b)**.** To further increase the efficiency of image processing, we introduce an optional pre-filtering step based on a pose normalization approach, which provides coarse subtomogram alignments (Supplementary Note 1.2) and efficient coarse classifications using k-means clustering. The coarse cluster averages provide rough shape information, which can be used to exclude groups of particles of low interest. For examples, clusters whose averages appear to contain only gold beads or consist only of membranes could be discarded, along with their subtomograms, to save computation time. On the other hand, clusters whose shapes resemble large, globular complexes can be merged and further processed using MPP. In addition, structures with very distinct shapes and sizes can be separated and processed separately to substantially save processing time. This preprocessing step is particularly useful for analysis of whole cellular tomograms, which have a large number of complexes that vary widely in shape, size and abundance. In practice, a pre-filtering procedure can efficiently reduce a collection of tens of thousands of automatically selected subtomograms to sets of a few thousand subtomograms, to be processed independently with MPP.

# Results

To assess the performance of our method, we applied it to three experimental cellular tomograms from different bacteria species, and carried out two types of studies using benchmarks of realistically simulated tomograms.

**Complexes under low crowding conditions.** First, we assessed MPP with simulated subtomograms as they would be extracted under relatively low crowding conditions such as relatively thin samples of purified macromolecular complexes or cell extracts. We applied our method to simulate 11,230 realistically and strongly distorted subtomograms, containing a benchmark mixture of 22 different macromolecular complexes with a wide range of abundances (Supplementary Note 2.1.3). To our knowledge, this is a substantially larger number of subtomograms and structures than any previously published classification study. The benchmark set of complexes is selected to have a wide range of variation in size and abundance, as well as shapes (Figure 2b).

After 32 iterations our results converged and MPP detected 12 patterns from the highly distorted subtomograms (Figure 2ab and Supplementary Data 1). In general, subtomograms of a complex were highly abundant in no more than one pattern (Figure 2a). Also the detected patterns were highly homogenous with respect to subtomogram membership. All twelve patterns were enriched with one dominant complex and the shapes of all detected pattern averages were very similar to the corresponding true complexes. Eight patterns uniquely matched complexes with a false discovery rate (FDR) ≤ 10%. Among these, four patterns had a FDR of 0%, meaning that all the subtomograms in each pattern were from the same true class. Also, the structural consistency (in terms of FSC, with 0.5 cutoff) between the average densities of the 8 complexes and the corresponding ground truth structures is very high: it ranges between 4.7 nm to 5.3 nm, which is comparable to the applied resolution. The best performances were achieved with the largest complexes, such as Glutamine Sythetase (2GLS) (FDR=0%, 88% of subtomograms detected), GroEL (1KP8) (FDR=0%, 75% of subtomograms detected), 50S ribosomal subunit (2AWB) (FDR=0%, 57% of subtomograms detected) and 20S Proteasome (3DY4) (FDR=0%, 48% of subtomograms detected), Carbomyl Phosphate synthase 1BXR) (FDR=1%, 42% of subtomograms detected), RecA hexamer (FDR=6%, 35% of subtomograms detected). Three patterns had larger FDRs (21%, 30% and 45%), however in each of these patterns essentially only a single complex was falsely co-assigned, which had very similar shapes to the target complex at the given resolution, which explains why the overall shape of the target complex was still very well predicted.

Seven complexes were not recovered (PDB ID: 1F1B, 1GYT, 1VPX, 2H12, 2IDB, 2GHO, 1QO1). The majority of these had relatively low abundance (< 200 instances), relatively small size and possibly non-discriminative shape features. Importantly, following MPP's design strategy, the subtomograms of these complexes were *not* wrongly assigned to any pattern but were simply left out, emphasizing the importance of the pattern mining approach in detecting high-quality structural patterns rather than attempting to classify all the subtomograms. All our results are highly reproducible. When we repeated our analysis with different random initial orientations for all the subtomograms the same complexes could be detected with similar FDR ranges (Supplementary Data 2). We also repeated our analysis with different random abundances for the complexes (Supplementary Data 3). With larger copy numbers, now two additional large complexes that previously had remained undetected were very well recovered: Aminopeptidase A (1GYT: FDR=1%, 79% of subtomograms detected) and Transaldolase (1VPX: FDR=0%, 48% of subtomograms detected). Two relatively small complexes could still not be detected even though they had larger copy numbers, which indicates that their relatively small size prevented their reliable detection at the given resolution. Our analysis suggests that a minimum copy number of 200-300 instances are necessary so that complexes can be reliably detected at the given resolution.

**Complexes under high crowding conditions.** Next, we tested MPP on realistically simulated cryo-electron tomograms of environments similar to those found in cell cytoplasm, which contained crowded mixtures of macromolecular complexes from the same benchmark set of 22 complexes used in the previous experiment (Supplementary Note 2.1.4.1). The crowding level of the simulated tomogram is 15.2%, which falls into the expected range for a cell[19–22] (Figure 2d, Supplementary Figure 8). The distortion level of the simulated tomogram is similar to that observed in cellular tomograms (Supplementary Note 2.1.4.3.1). We used automated ‘Difference of Gaussian’ particle picking[23] without using a structural template to extract subtomograms likely to contain one target complex (Supplementary Note 2.1.4.2). Particle picking generally favors larger complexes and 11 out of the 22 complexes were extracted with at least 200 instances (11 smaller complexes had fewer than 140 extracted subtomograms) (Figure 2c). Extracted subtomograms may also contain fragments of neighboring structures. We therefore applied our method for automatically masking target complexes at each MPP iteration (Step 8b) (Supplementary Note 1.1.7). This test case is substantially more challenging than the previous one, because errors in automated particle picking and target complex segmentation can influence the MPP performance. Despite these challenges, MPP detected six patterns, four of which with false discovery rates ≤ 23% and very well predicted shapes with structural consistencies ranging from 4.3 nm to 4.8 nm and (Figure 2cd, Supplementary Data 4, and Supplementary Note 2.1.4.3). Among these, one (50S ribosome / 2AWB) had a false discovery rate of 0%. MPP also predicted two patterns that are a mix of complexes. These two have structural consistencies ≤ 6.5nm and are very similar in shape to the most abundant complex in the pattern. One of these patterns (pattern 5 - 2GLS) contained only two complexes of relatively similar shapes. The other detected pattern with the smallest size is a mixture of more than 10 complexes. Most of these complexes have relatively low abundance after particle picking and are of relatively small and similar shape as evidenced by their tight clustering based on shape similarity when comparing all target complexes (Supplementary Figure 10 and Supplementary Data 4a). At the given resolution and crowding level it is not possible to distinguish these small complexes. However, MPP still predicted their size and location. To test the reproducibility, we repeated our analysis by simulating another crowded tomogram with random positions and orientations of the complexes. Now even 6 patterns were successfully recovered at FDR < 30%. Two additional complexes were detected, mainly as a result of their increased copy numbers after particle picking. (Supplementary Note 2.1.4.4.1, Supplementary Data 5, Supplementary Figure 11). We also tested MPP with simulated tomograms with a lower crowding level (Supplementary Figure 12). In this case MPP gives improved performance and detects a total of 9 patterns (Supplementary Note 2.1.4.4.2, Supplementary Data 6, and Supplementary Figure 13).

**Experimental cellular tomograms.** We tested MPP on three cellular cryo-electron tomograms of bacteria, namely single cells of lysed *Acetonema longum*, intact *Hylemonella gracilis* sample and intact *Bdellovibrio bacteriovorus* sample. We performed automated, template-free particle picking to extract a total of ~30,000 subtomograms from the three cells as described in Supplementary Notes 2.2.1, 2.2.2, and 2.2.3. For intact cells (*H. gracilis* and *B. bacteriovorus*) the cell regions are manually segmented and only the subtomograms within the cells are extracted. The cell of *A. longum* appeared lysed and particles were noticeable also at the cell exterior, which was included in the analysis. We then applied pre-processing (Supplementary Note 1.2) and applied MPP separately for each cell type.

For the three cells, MPP discovered 12, 15 and 10 patterns of relatively high quality score for *A. longum*, *H. gracilis* and *B. bacteriovorus*, respectively (Figure 3ab; Supplementary Notes 2.2.1.3, 2.2.2, and 2.2.3; Supplementary Data 7, 8 and 9; Supplementary Movies 1, 2 and 3): the resolution of these patterns (in terms of Gold Standard FSC) ranged from 4.1-5.8 nm, from 3.5-10.5 nm and from 4.8-15.0 nm respectively. These ranges reflect different degrees of reproducibility for the patterns of a given cell type. The shapes and positions of some patterns already give indications as to the identity of the complexes. For example, several different patterns clearly represent membrane particles lining the cell boundaries (Figure 3ab) (e.g., patterns 2, 3 and 7 for *B. bacetriovorus*, Supplementary Figure 22 and Supplementary Data 9b). Other patterns are small globular complexes of various different shapes (Supplementary Data 7b, 8b and 9b) (e.g., patterns 4, 5, 6, 7, 8, 10, and 12 for *H. gracilis*, Supplementary Figure 19). Some larger patterns have shapes and sizes very similar to the structures of GroEL (pattern 4 in *A. longum*) and ribosome (patterns 0, 1, 2 in *H. gracilis*; and patterns 0, 1, 9 in *B. bacteriovorus*), and were also observed at relatively large abundance (e.g. a total of 802 copies of ribosome-like patterns in *H. gracilis*). We refined these patterns further using the GA method (Supplementary Note 1.1.3.3). Figure 3c demonstrates the similarity between these structures and GroEL and 70S ribosome templates simulated from their atomic structures.

Strikingly, when we fit the atomic structure of GroEL into the average density of pattern 4 in *A. longum* we observed a remarkably good fit: the size and shape of pattern 4 is in very good agreement with the GroEL X-ray structure (Figure 3c, Supplementary Figure 16). We further assessed the likelihood of pattern 4 being a GroEL analog by several criteria, using a template-based search (Supplementary Notes 2.2.1.4, 2.2.2 and 2.2.3). Specifically, we aligned all the subtomograms from each cell type against a collection of the 28 different template structures most abundant in cells. We found that the alignment scores for subtomograms of the GroEL-like pattern 4 (resolution 4.5 nm, Supplementary Data 7b) were statistically significantly higher to the GroEL template (PDB ID: 1KP8) than to any other template (one-sided Wilcoxon rank-sum test with p-value $3.2^{-10}$, Supplementary Figure 17a), confirming the clear visual similarity of pattern 4 with GroEL. The second closest match was the GroEL/GroES complex (PDB ID: 1AON), although this template had significantly lower alignment scores. Also, we showed that the subtomograms of pattern 4 had the strongest matches to the GroEL template, in terms of alignment scores, compared to the rest extracted subtomograms of *A. longum* (p-value $< 2.2 \times 10^{-220}$, Supplementary Figure 17b). These tests indicate that our template-free approach yields similar results to a template matching approach with GroEL as a template structure. All these observations support the hypothesis that the subtomograms in pattern 4 contain a bacterial analog of the GroEL complex. Interestingly, the high abundance of GroEL complexes (481 instances) is observed only in the *A. longum* cell and may be related to a stress response. We note that this cell appeared to be dead and lysed before image acquisition[24].

Equally convincing are the assessments of ribosome structures in *H. gracilis* and *B. bacteriovorus* cells. In *H. gracilis*, three patterns (patterns 0, 1, and 2) are visually similar to ribosome structures (Figure 3c). This observation is confirmed by template-based assessments. The subtomograms in pattern 0, 1, and 2 had indeed the highest alignment scores with the ribosome template (both the full ribosome PDB ID: 2J00-2J01 and its 50S subunit with PDB ID: 2AWB) (Supplementary Figure 20c) (p-value $< 4.1 \times 10^{-22}$) compared to any of the other 26 templates showing that all three patterns are most likely ribosome structures. Subtomograms of pattern 1 (resolution 8.7nm, Supplementary Data 8b) had the highest alignment scores (p-value $< 1.5 \times 10^{-65}$, Supplementary Figure 20a) with the ribosome templates. Also when comparing all the rest extracted subtomograms, those in pattern 1 had significantly higher alignment scores with the ribosome templates (p-value $< 2.0 \times 10^{-125}$, Supplementary Figure 20b). All these observations support the hypothesis that these patterns contain a ribosome structure.

Similarly, in *B. bacteriovorus*, the subtomograms in pattern 1 (resolution 12.0 nm, Supplementary Data 9b) were visually similar to the ribosome and had significantly higher alignment scores to the ribosome templates (PDB ID: 2J00-2J01 and 50S subunit with PDB ID: 2AWB) compared to any of the other 26 templates (p-value $< 1.7 \times 10^{-24}$, Supplementary Figure 23a). Compared to all detected patterns, subtomograms of pattern 1 had the highest alignment scores to the ribosome template (PDB ID: 2J00-2J01) (Supplementary Figure 23c) and also had the highest-ranking scores compared to all other extracted subtomograms (p-value $< 6.3 \times 10^{-6}$, Supplementary Figure 23b). With similar shapes, subtomograms from patterns 0 and 9 also have relatively high alignment scores with respect to the ribosome template and also likely contain a ribosome complex.

Interestingly, we found distinct spatial distributions for the different complexes in the *B. bacteriovorus* tomogram. For instance, the likely ribosomal patterns are excluded from a region along the central axis of the cell (Supplementary Figure 24b), where the bacterial nucleolid is expected to be. It is likely that ribosomes would be located close to, but not directly overlapping with, regions of the genome. Ribosome-like structures also appear to be less abundant in the tip region associated with the bacterial flagella motor, although we cannot exclude the possibility of imaging artifacts being partially responsible for the lack of ribosome structures in this region. A smaller pattern was only enriched in the tip of the bacteria (patterns 4 and 5, Supplementary Figure 24c). Interestingly, a different small pattern was found only in the region that occupies the bacterial nucleolid genome (pattern 6) (Supplementary Figure 24d).

## Summary

In summary, our MPP method is designed to analyze in a high-throughput and template-free fashion a large number of subtomograms containing many structural classes, and derive robust structural patterns. More importantly, our method represents a substantial step towards visual proteomics analysis inside single cells. Automatic pattern mining inside cellular ECT tomograms is still very challenging and our approach is only a first step in this direction. On the other hand, together with recent breakthroughs on direct detectors[25] and phase plates[26], which significantly improve contrast and resolution of cellular ECT data, correlative light and electron microscopy[27], and focused ion beam milling[28], which enables ECT to image a substantially larger variety of cell types, we expect that our method can become an integral part of visual proteomics applications. In addition, MPP is also useful for analyzing tomograms containing isolated but highly heterogenic particle mixtures, such as cell lysates. Moreover, once patterns are detected in a whole cell analysis, they can be used by other methods such as template searches [e.g. [2,5,7,29]], subtomogram classifications [e.g. [10,12,16,30–32]] and subtomogram averaging methods [e.g. [9,33]] for further refinement. Therefore, our work complements existing template-based and template-free methods.

## Acknowledgements

This work was supported by NIH R01GM096089, Arnold and Mabel Beckman Foundation (BYI), NSF career 1150287 to F.A.. Funding from the Howard Hughes Medical Institute to G.J.J.

# Figures

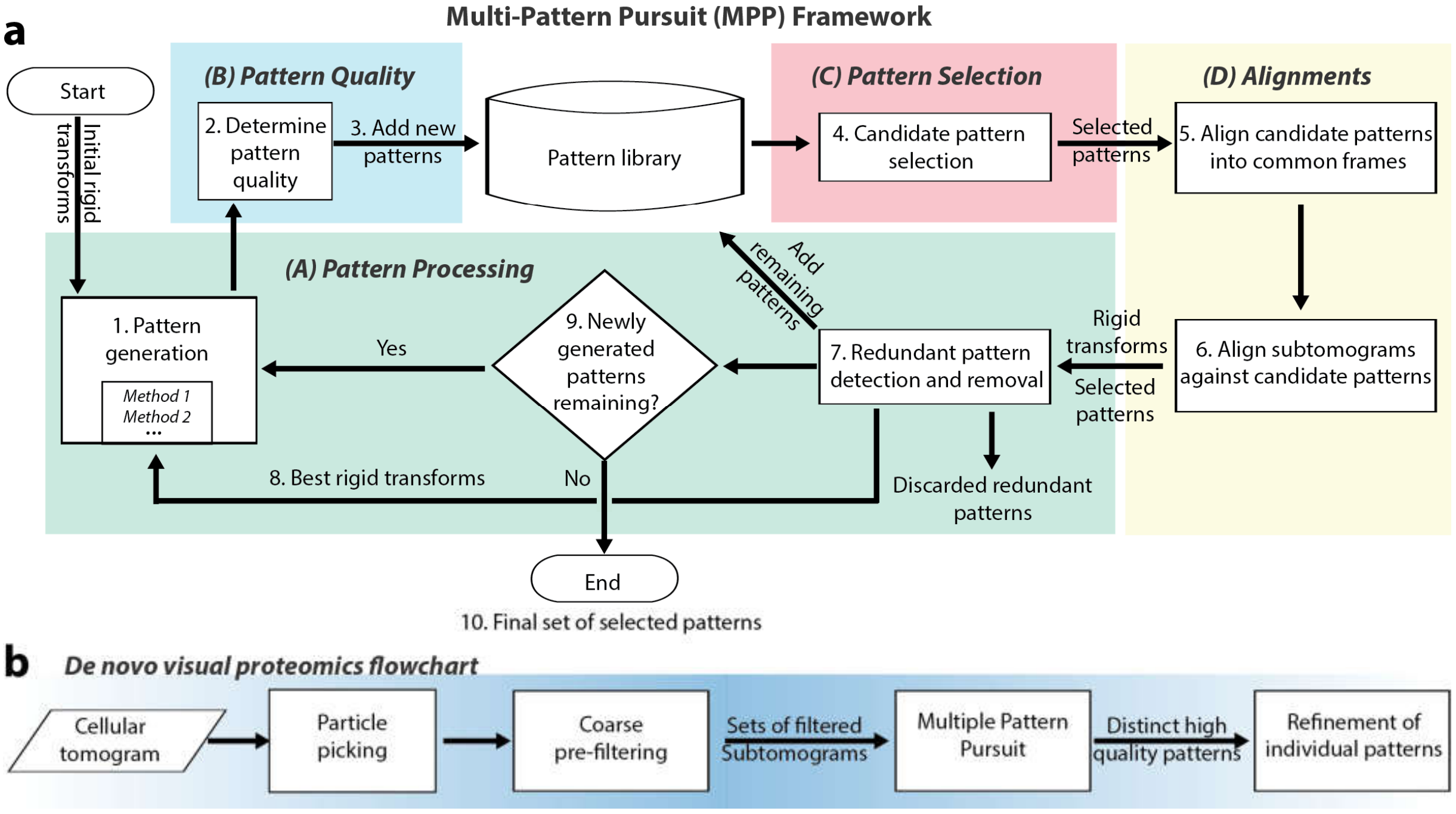

**Figure 1**: Overview of the method. (a) Flow chart of the MPP framework (Supplementary Note 1.1.1). In the flow charts, actions are in boxes, data are on arrows, and diamonds represent decisions. (b) Flow chart of overall processing pipeline, including preprocessing and postprocessing of cryo-electron tomograms. The pre-processing step consists of pose-normalization based coarse alignments of the subtomograms, combined with an initial filtering through k-means clustering to define sets of subtomograms containing similarly sized particles.

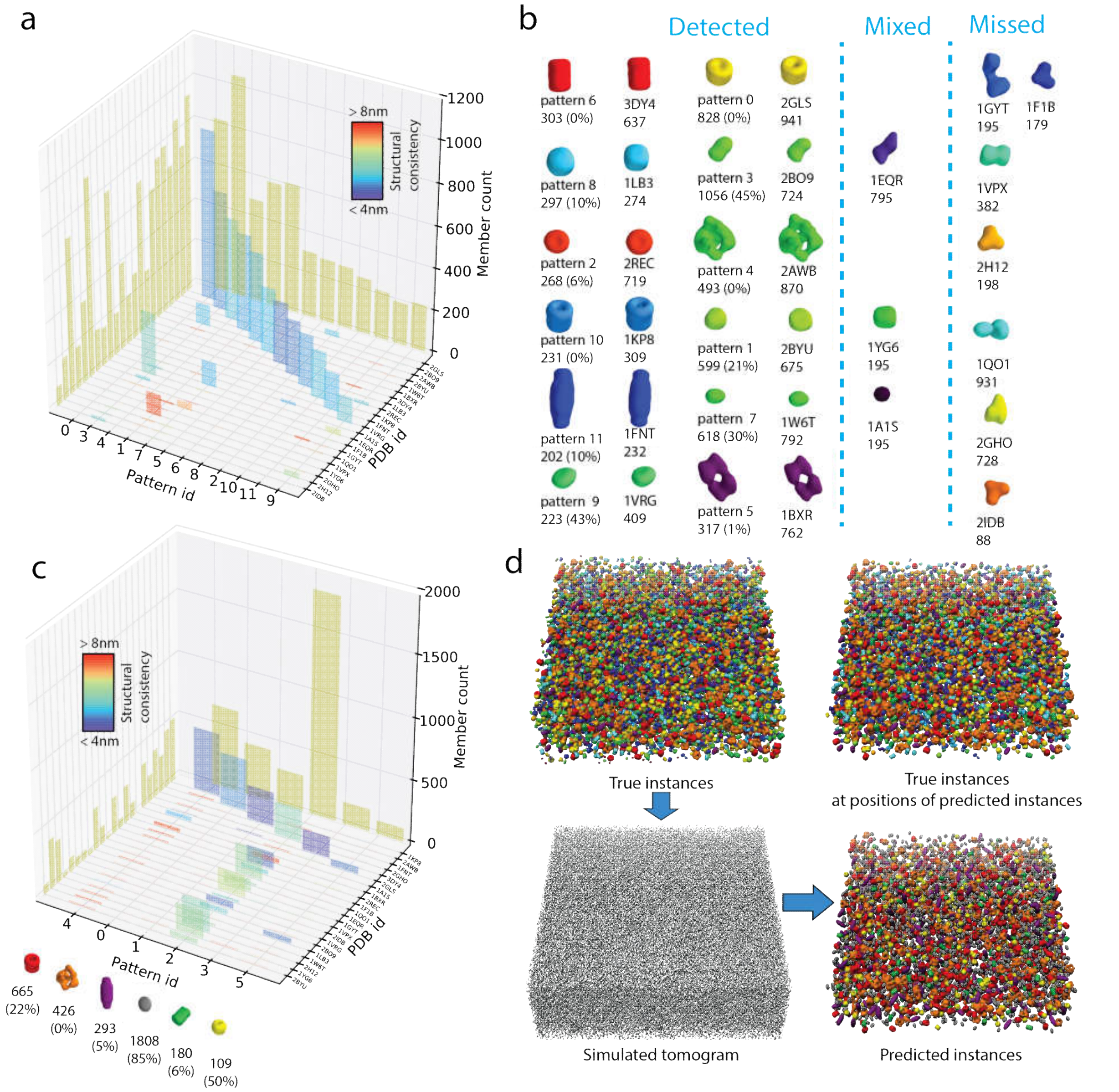

**Figure 2**: Comparison of MPP pattern mining results with ground truth complexes for individually simulated subtomograms and tomograms containing a crowded mixture of complexes. (a) Results of individually simulated subtomograms, column plot representation of the contingency table (Supplementary Data 1c) of the subtomogram membership overlap between true and inferred patterns. The height of each column at each axis corresponds to the total number of subtomograms of the ground truth complex and the total number of subtomograms in the predicted patterns, respectively. The height of each column inside the table corresponds to the number of subtomograms for each ground truth complex in each predicted pattern. The colors of the columns indicate the structural consistency between the ground truth and the corresponding pattern averages, quantified as FSC with cutoff 0.5 (Supplementary Section 2.1.2). (b) The isosurfaces of predicted patterns compared to ground truth structures. The ground truth structures are indicated by their PDB ID code, and the number of instances. Also shown are the isosurface representations of the predicted patterns with the number of instances and the false discovery rate (FDR) in parentheses. The FDR shows the fraction of wrongly assigned subtomograms in the pattern. (c) Column plot representation of the contingency table for the simulated cellular tomogram of a crowded mixture of complexes (Supplementary Data 4c). Also shown are the isosurface representations of the predicted patterns with the number of instances and FDR in parentheses. (d) Upper left panel: isosurface of the ground truth mixture of crowded complexes. Lower left panel: isosurface of the simulated tomogram. Lower right panel: isosurface representation of the predicted patterns and their localizations. Upper right panel: True instances that were detected using MPP.

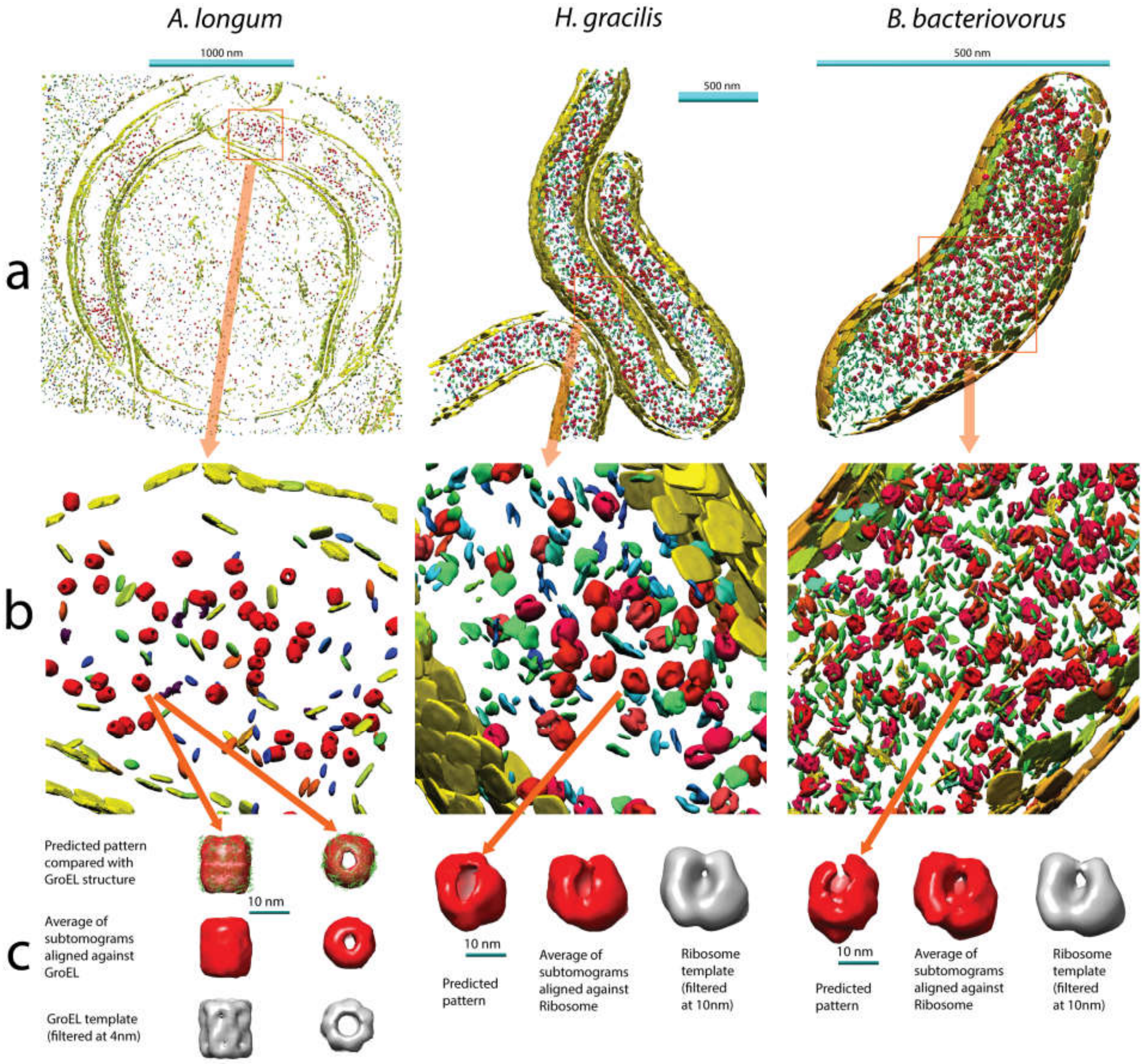

**Figure 3**: Discovered patterns in three cellular tomograms: *A. longum*, intact *H. gracilis*, and intact *B. bacteriovorus* cells (left, middle, right). (a) Embedded instances of patterns. (b) Embedded instances, zooming in on a particular region. (c) Isosurfaces of one example pattern from each experiment. The GroEL-like pattern is also fitted with a known atomic model of GroEL. Isosurface of the average density of the example pattern, aligned with the known structures of the GroEL (PDB ID: 1KP8) and ribosome complexes (PDB ID: 2J00-2J01).

# *De novo* visual proteomics of single cells through pattern mining – supplementary notes and figures

Min Xu[1], Elitza I Tocheva[2], Yi-Wei Chang[3], Grant J Jensen[3,4], Frank Alber[1*]

[1] Program in Molecular and Computational Biology, University of Southern California, Los Angeles, California, USA.

[2] Department of Stomatology, Universit de Montral, Montreal, Quebec, CANADA.

[3] Division of Biology and Biological Engineering, California Institute of Technology, Pasadena, California, USA.

[4] Howard Hughes Medical Institute, Pasadena, California, USA.

# Contents

*Corresponding author

# List of Figures

# 1 Methods

## 1.1 Multiple Pattern Pursuit (MPP) framework

### 1.1.1 Main framework

Multiple Pattern Pursuit (MPP) takes a collection of subtomograms and searches for structural patterns. A structural pattern is defined as a set of rigidly transformed subtomograms and their density average . These subtomograms are similar to each other and are likely to contain the same structure. MPP aims to maximize the quality (in terms of SFSC score, Supplementary Note 1.1.2) of multiple distinct patterns extracted from these subtomograms. It is an iterative optimization process that searches for patterns in the pattern space. Such space is the Cartesian product of pattern membership and rigid transform of subtomograms. MPP combines novel components and our previously developed components. Each iteration of MPP consists of following steps (**Figure 1a** in the main text):

1. Based on current rigid transformations $T$ of subtomograms, generate a collection of candidate patterns $S^{\text{candidate}}$ (Supplementary Note 1.1.3).
2. Determine quality of the patterns in $S^{\text{candidate}}$ in terms of their SFSC scores (Supplementary Note 1.1.2).
3. Add $S^{\text{candidate}}$ into the pattern library $L$: $L \leftarrow L \cup S^{\text{candidate}}$
4. Select a set $S^{\text{sel}}$ of high quality patterns from $L$ under the constrain of minimal subtomogram membership overlap (Supplementary Note 1.1.4).
5. Align the subtomogram averages of patterns in $S^{\text{sel}}$ into common reference frames (Supplementary Note 1.1.5).
6. Align all subtomograms against each of the subtomogram averages of all patterns in $S^{\text{sel}}$.
7. Identify structurally redundant patterns $S^{\text{redundant}}$ in $S^{\text{sel}}$ (Supplementary Note 1.1.6). Remove patterns in $S^{\text{redundant}}$ from $L$: $L \leftarrow L \setminus S^{\text{redundant}}$. In other words, patterns in $S^{\text{redundant}}$ will never be selected in future iterations.
8. Update subtomogram transformations $T$ according to the best alignment between the subtomograms and the subtomogram averages of the remaining selected patterns in $S^{\text{remain}} := S^{\text{sel}} \setminus S^{\text{redundant}}$
9. If the patterns in $S^{\text{remain}}$ are all generated from at least $n^{\text{stop}}$ iterations earlier, stop. Otherwise, continue to next iteration.

Remarks:

- For high-throughput processing, we use our fast alignment method [99]. Alternative alignment methods [e.g. 32, 5, 98, 101, 19, 105] may also be used. Alignment methods could fail when the SNR of tomograms is very low.
- Empirically, we set $n^{\text{stop}} = 5$, which we found is sufficiently large to minimize the chance of missing new and even higher quality patterns. A slight variation of $n^{\text{stop}}$ does not change the outcome of our analysis.
- The software implementation of MPP is based on a variant of the TomoMiner platform [36].

### 1.1.2 3D SSNR based FSC that take into account missing wedge effect

In pattern mining, a measure of quality of the subtomogram average is needed for the optimization process. Following the common practice in the cryo-electron microscopy (CEM) and cryo-electron tomography (ECT) fields, we measure the quality of a subtomogram average by the level of structural details of the pattern that the average can confidently represent, i.e. the resolution of the average, which is widely used for validating subtomogram averages. Such resolution is often calculated through measuring relative uncertainty or reproducibility. There are two main types of such measures [50]: The first type of measure is the Spatial Signal to Noise Ratio (SSNR ) [92, 74], which compares homogeneous structural signal against structural and non-structural variations. The second type of measure, the Fourier Shell Correlation (FSC) [83, 93], is a measure on reproducibility. FSC is calculated by randomly splitting the set of subtomograms into two halves and by measuring the consistency (at different scales) between the corresponding two averages from the two halves. FSC has different variants [50].

We use a SSNR based FSC score as the quality measure. There are several advantages to use such combinatio n (compared to calculating FSC from splitting the data into two halves). First, SSNR is directly computed from all subtomograms and therefore it reduces the underestimation of the resolution due to the sample size limit, and there is no uncertainty introduced by the statistical fluctuation from the random choice of splitting [50]. Second, the measure can be efficiently computed in parallel, enabling high-throughput processing (see Supplementary Note 1.1.2.1). Third, SSNR can be easily extended to take into account missing wedge effects, which is one of the major distortions in the ECT imaging process. On the other hand, our experience shows that the use of SSNR alone as a quality measure may not be sufficient. It has an undesired property: its tends to emphasize on low frequency components because the SSNR measure ranges from zero to infinity, and its value decreases dramatically as the frequency increases. Therefore it would be beneficial to use a normalized measure like FSC that accounts more high frequency information. To our knowledge, the subtomogram average quality measure has not been used as objective in any existing template-free subtomogram classification methods.

Formally, we denote a set of $n$ aligned subtomograms as $f_1, \ldots, f_n$, their Fourier transform as $F_1, \ldots, F_n$, and the corresponding wedge masks as $M_1, \ldots, M_n$ (as defined in Equation 3). We adapt the standard SSNR measure to take into account the missing wedge effect and derive a SSNR measure $\eta_r$ at frequency $r$:

$$\eta_r = \frac{\int_{||\xi|-r|\leq \Delta r} \tilde{M}(\xi)|\mu(\xi)|^2}{\int_{||\xi|-r|\leq \Delta r} \sigma^2(\xi)} \tag{1}$$

where $\Delta r = 1$, $\tilde{M}$ is the summation of the missing wedge masks:

$$\tilde{M}(\xi) := \sum_i M_i(\xi)$$

,

$$\mu(\xi) = \frac{\sum_i M_i(\xi) F_i(\xi)}{\tilde{M}(\xi)}$$

and

$$\sigma^2(\xi) = \frac{\sum_i M_i(\xi)|M_i(\xi)F_i(\xi) - \mu(\xi)|^2}{\tilde{M}(\xi) - 1}$$

Given the above calculated SSNR the FSC $\rho_r$ at frequency $r$ can be estimated according to [35, 50]:

$$\rho_r = \frac{\eta_r}{2 + \eta_r}$$

We use the sum of FSC over all frequencies (denoted as *SFSC*) to score the quality of a subtomogram average:

$$\tilde{\rho} := \sum_r \rho_r \quad (2)$$

The higher the $\tilde{\rho}$, the higher is the quality of the corresponding subtomogram average of a pattern.

**1.1.2.1 Additive property** The calculation of FSC can be easily parallelized due to the following property: $\eta_r$ can be calculated from $\tilde{M}$, $\sum_i M_i F_i$, and $\sum_i F_i \overline{F_i}$, where $\overline{F_i}$ is the complex conjugate of $F_i$. All these three quantities are additive for disjoint sets of subtomograms.

### 1.1.3 Candidate pattern generation

The MPP optimization is performed in two stages, which differ in the way candidate patterns are generated. After stage 1 terminates, the MPP starts stage 2 with the rigid transforms $T$ and pattern library $L$ resulted from stage 1.

The main purpose of stage 1 is to obtain updated $T$ so that subtomograms with the same underlying structures are roughly aligned, and obtain a first estimate of (the number of) distinct patterns. Stage 1 begins with an initially empty pattern library $L$ and randomly assigned rigid transforms $T$ for all subtomograms, which are updated at the end of every iteration. In stage 1, the pattern generation is performed by a dimension reduction approach (Supplementary Note 1.1.3.1) followed by k-means clustering with a fixed cluster number $k^{\text{k_means_fix}}$, which is usually chosen to over-partition the collection of subtomograms. When the true set of structurally distinct patterns is unknown, an intuitive strategy is to over-partition the number of clusters then identify and remove the clusters leading redundant patterns (see Supplementary Note 1.1.6).

After stage 1 terminates, the MPP starts stage 2 with the $T$ and $L$ resulted from stage 1. In stage 2 the subtomogram membership and density averages of the patterns are improved. In stage 2, two independent methods are used to generate candidate patterns (resulting patterns of both methods are added to the pattern library): First, the sequential expansion method (Supplementary Note 1.1.3.2) and second, dimension reduction (Supplementary Note 1.1.3.1) followed by k-means clustering in which the cluster number $k^{\text{k_means_adaptive}}$ is assigned adaptively according to $|S^{\text{remain}}|$ of the last iteration: $k^{\text{k_means_adaptive}} \approx k^{\text{k_means_adaptive_factor}} |S^{\text{remain}}|$, where $k^{\text{k_means_adaptive_factor}} = 1.2$ is a fixed ratio.

Remarks: Each time, the k-means clustering is repeated 10 times and the best clustering result is chosen in order to reduce the chance to be trapped in a local minima. We use the k-means++ initialization [2, 21] to improve convergence. Such procedure has been implemented in the off the shelf sklearn package [72].

**1.1.3.1 Imputation based dimension reduction** Dimension reduction for high dimension data has been extensively studied in different areas [30, 15]. It is very useful for extracting key low dimension features that contain the majority of discriminative signals across images, and reducing the influence of non-informative variance. Dimension reduction is also very useful for significantly speeding up clustering. This is because subtomograms are high dimension data, and computation of distances between two subvolumes in a smaller number of dimensions is much more computationally effective than direct calculating distances in their original high dimensional space.

One major obstacle for directly applying existing dimension reduction methods is the missing wedge effect due to the missing as a result of the limited tilt angle range of captured projection images. As a result, the objects in a subtomogram have anisotropic resolutions across different directions, which introduces bias to the dimension reduction [32, 5]. The missing wedge effect can be described in Fourier space, where the Fourier coefficients in certain regions are missing. The locations of Fourier coefficients $\mathcal{F}f$ with valid values and missing values can be represented using a missing wedge mask function $M$.

$$M(\xi) := \begin{cases} 1 & \text{if the Fourier coefficient at } \xi \text{ is valid} \\ 0 & \text{if the Fourier coefficient at } \xi \text{ is missing} \end{cases} \quad (3)$$

where $f : \mathbb{R}^3 \to R$ is the function that represents image intensity of a subtomogram; $\mathcal{F}$ is the Fourier transform operator; and $\xi \in \mathbb{R}^3$ is a location in the Fourier space.

Two typical types of strategies have been proposed to handle the missing wedge effect for dimension reduction. The first type omits the Fourier coefficients that are not used for dimension reduction [e.g. 41]. The second type estimates missing values [e.g 104]. These methods are effective for enhancing the subtle true discriminative signal across aligned subtomograms. However, these methods are generally designed for cases in which the underlying structures of all subtomograms are similar to a single reference density map, which does not apply to a Visual Proteomics setting with the existence of high degree of structural heterogeneity among subtomograms.

To solve this problem, we propose an imputation strategy. We use for each subtomogram its currently best aligned density map (chosen from the set of pattern density maps in $S^{\text{remain}}$ obtained from the last iteration of MPP(Supplementary Note in 1.1.1)) as a reference to replace the missing Fourier coefficient values with those from the density map, Supplementary Figure 1 illustrates the basic idea.

Formally, we want to use Fourier coefficients of a reference density map $a$ as an estimate of the missing Fourier coefficients of a subtomogram $f$, given that aligning $f$ against $a$ gives the best alignment score compared to aligning $f$ against other maps in the same collection. For simplicity, suppose $f$ has been rigid transformed according to its alignment against $a$, and $M$ be the corresponding missing wedge mask of $f$ rotated according to the rigid transform. Then we can form a transformed and imputed subtomogram $\tilde{f}$ such that:

$$(\mathcal{F}\tilde{f})(\xi) := \begin{cases} (\mathcal{F}f)(\xi) & \text{if } M(\xi) = 1 \\ (\mathcal{F}a)(\xi) & \text{if } M(\xi) = 0 \end{cases} \tag{4}$$

After imputation, in principle any generic dimension reduction method can be directly applied without any modification to take into account missing wedge effects. After dimension reduction, in principle the consequent clustering step also does not need to take account of missing wedge effects.

After imputation, to speed up processing in the dimension reduction, we combine feature selection and feature extraction. We first calculate the average covariance between neighbor voxels in a similar way as our previous work [99]. We then select a number (usually 10,000) of voxels with highest and positive average covariance (feature selection step) and apply EM-PCA [79, 4] (feature extraction step) to perform dimension reduction. When the extracted dimension number is relatively small, EM-PCA can be very fast, scalable [4] and memory-efficient compared to other Principal Component Analysis (PCA) methods. It can normally handle tens of thousands of subtomograms using a single CPU core in a couple of hours. Empirically, we found a dimension number of 50 to be able to capture sufficient data variance for clustering the subtomograms.

Remarks: When using the imputation based dimension reduction for MPP (Supplementary Note 1.1.1), all subtomograms are first imputed. The calculation of the principal directions of PCA is done only using subtomograms of the non-redundant selected patterns $S^{\text{remain}}$ obtained from the last iteration . Then we project all imputed subtomograms onto the principal directions.

#### 1.1.3.1.1 Proof of equivalence between wedge-masked difference and imputed difference

The difference between $a$ and $\tilde{f}$ can be treated as a generalization of the wedge-masked difference proposed in [41], where the wedge-masked difference is equivalent to a special case of our approach where only a single average is used to impute all the aligned subtomograms and calculate differences among these subtomograms.

Without band limit, the wedge-masked difference [41] between a reference density map $a$ and an aligned subtomogram $f$ (with corresponding rotated wedge mask $M$) is calculated as

$$\begin{aligned} &\mathcal{F}^{-1}[M(\mathcal{F}a)] - \mathcal{F}^{-1}[M(\mathcal{F}f)] \\ &= \mathcal{F}^{-1}[M(\mathcal{F}a - \mathcal{F}f)] \end{aligned}$$

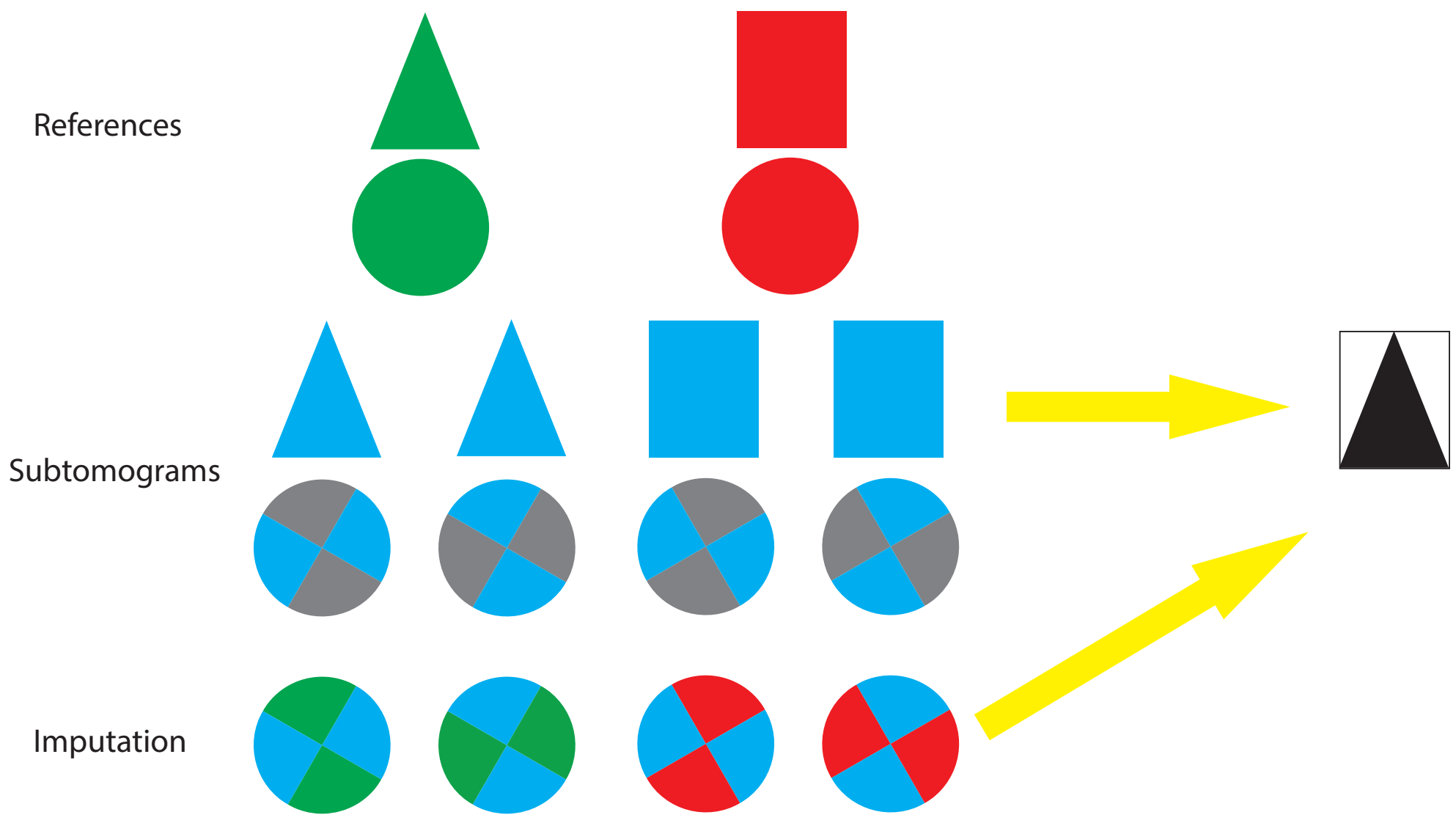

Supplementary Figure 1: The basic idea of imputation based dimension reduction. Left: Upper row: Green triangle and red rectangle are structures in two reference density maps. The filled circles show missing wedge masks, which are regions of valid Fourier components of the subtomogram image in Fourrier space. Middle row: Triangles and rectangles are structures in individual subtomograms. These subtomograms are aligned against its most similar references (top row). The circles are the corresponding missing wedge masks, which indicate regions with valid (colored in blue) and missing (colored in grey) Fourier coefficients. Lower row: Imputation of subtomograms by replacing missing Fourier coefficient regions (previously in grey color) with valid Fourrier coefficients from the corresponding references (in green and red colors). Right: The variance of voxel intensities across imputed subtomograms. The region with low variance is represented in black color. The region with high variance is represented in white color. The use of variance map is only for illustration purpose.

. According to Equation 4,

$$
\begin{aligned}
&[M(\mathcal{F}a - \mathcal{F}f)](\xi) \\
&= \begin{cases} (\mathcal{F}a)(\xi) - (\mathcal{F}f)(\xi) & \text{if } M(\xi) = 1 \\ 0 & \text{if } M(\xi) = 0 \end{cases} \\
&= \begin{cases} (\mathcal{F}a)(\xi) - (\mathcal{F}\tilde{f})(\xi) & \text{if } M(\xi) = 1 \\ (\mathcal{F}a)(\xi) - (\mathcal{F}\tilde{f})(\xi) & \text{if } M(\xi) = 0 \end{cases} \\
&= (\mathcal{F}a)(\xi) - (\mathcal{F}\tilde{f})(\xi)
\end{aligned}
$$

. Therefore

$$
\begin{aligned}
&\mathcal{F}^{-1}[M(\mathcal{F}a)] - \mathcal{F}^{-1}[M(\mathcal{F}f)] \\
&= \mathcal{F}^{-1}[\mathcal{F}a - \mathcal{F}\tilde{f}] \\
&= a - \tilde{f}
\end{aligned}
$$

**1.1.3.2 Sequential expansion** Besides using k-means clustering, we also use sequential expansion as a heuristic for generating candidate patterns. Sequential expansion adds those subtomograms from a pattern if their inclusion increases the pattern quality. Therefore sequential expansion allows omission of subtomograms that are likely wrongly assigned to a pattern based on k-means clustering. All subtomgrams are ranked according to their alignment score to the pattern average. Then an alignment score cutoff is searched such that the quality of the pattern formed by the set subtomograms with scores higher than the cutoff maximizes the quality of the newly formed pattern average.

Formally, let $S^{\text{remain}}$ to be the non-redundant patterns selected from the last iteration of MPP (Supplementary Note 1.1.1). For each subtomogram average $a \in S^{\text{remain}}$, from all subtomograms we select those that have the highest alignment scores against $a$ compared to all other pattern averages in $S^{\text{remain}}$. Suppose in total there are $n_a$ such subtomograms, let $C = \{f_1, \ldots, f_{n_a}\}$ be the collection of subtomograms. They are aligned against $a$ and ordered in terms of alignment scores in descending order. Then, for each subcollection $\{f_1, \ldots, f_i\} \subset C, 1 < i \leq n_a$ of these subtomograms, we can calculate a SFSC score $\tilde{\rho}_i$ (Supplementary Note 1.1.2) of these subtomograms. Using the additive property (Supplementary Note 1.1.2.1), $\tilde{\rho}_{i+1}$ can be calculated efficiently from $\tilde{\rho}_i$ without re-scanning over $\{f_1, \ldots, f_i\}$. . Let $i^* = \arg\max_i \tilde{\rho}_i$, a new candidate pattern can be formed using $\{f_1, \ldots, f_{i^*}\}$. In such way, each pattern in $S^{\text{remain}}$ can be used to generate a new candidate pattern.

#### 1.1.3.3 Genetic algorithm

In MPP (Supplementary Note 1.1.1), the candidate patterns are generated by using k-means clustering and sequential expansion. After MPP iterations converged distinct patterns of highest SFSC scores are produced.

After MPP iterations converged to distinct set of patterns we also applied an optional refinement method to individual patterns to achieve even higher quality. We call such type of pattern mining as *Single Pattern Pursuit* (SPP). SPP assumes that the input collection of subtomograms is dominated by a single structure. Given a collection of subtomograms and their rigid transforms, we want to select a subset of subtomograms that maximizes the SFSC score defined in Equation 2 in Supplementary Note 1.1.2. Such optimization based subtomogram selection method does not require a manually specified cutoff to exclude non-homogeneous subtomograms. The optimization of this score is a nontrivial combinatorial optimization problem. We use Genetic Algorithm (GA) to perform such an optimization. Although such an approach is computationally intensive, it further improves the quality of a pattern with a small set (normally less than 1000) of subtomograms, usually on a single computer within a couple of hours.

A Genetic Algorithm (GA) [86] is a generic optimization technique that mimics the process of natural selection. Initially, the GA starts with a population of randomly generated candidate solutions. GA is an iterative process and the population of candidate solutions in each iteration is called a generation. In each generation, the fitness of every individual candidate solution is evaluated. The individual candidate solutions are randomly selected from the current generation with a probability that is proportional to the fitness of the solutions. The selected solutions are recombined and randomly mutated to form a new generation of candidate solutions.

In order to speed up the convergence, we follow the popular *elitism* heuristic [25, 80, 109] by keeping, besides a population of $n$ candidate solutions, also a population of $n$ top candidate solutions generated so far in previous iterations, and combine these two populations to generate a new generation of candidate solutions so that the top candidate solutions are carried over from one generation to the next unaltered.

In our implementation, given a set of $m$ subtomograms with fixed rigid transforms, we encode a candidate solution as a binary vector $\mathbf{o} \in \{0, 1\}^m$, which corresponds to a candidate pattern. Each element of $\mathbf{o}$ is 1 if the corresponding subtomogram is to be included into the corresponding candidate pattern, and 0 otherwise. Given any candidate solution, we can calculate a SFSC score of the average density of the corresponding selected subtomograms according to Equation 2. Such a score then represents the fitness of the corresponding candidate solution.

Our GA procedure is initiated by a population $O^0$ of $n$ randomly generated candidate solutions, and an empty pool $B^0 = \varnothing$ of top solutions. A particular iteration $i > 0$ consists of following steps.

1. Given a generation $O^{i-1}$ of the last iteration $i - 1$, calculate SFSC score for each candidate solution in $O^{i-1}$.

2. Use the combined population $C^{i-1} = B^{i-1} \cup O^{i-1}$ to form a generation $O^i$:
   (a) Randomly select a pair $P$ of candidate solutions in $C^{i-1}$
   (b) Perform crossover operation (Supplementary Figure 2) followed by mutation operation to generate a pair $P'$ of new candidate solutions
   (c) Add both candidate solutions $P'$ into the new population $O^i$.
   (d) Repeat the above steps until $|O^i| \geq n$
3. Combine $O^i$ and $B^{i-1}$ to form a new population of top candidate solutions $B^i$

The iterative process continues until the best candidate solution in $B$ is unchanged for a fixed number of iterations. It selects the best candidate solution as the final solution.

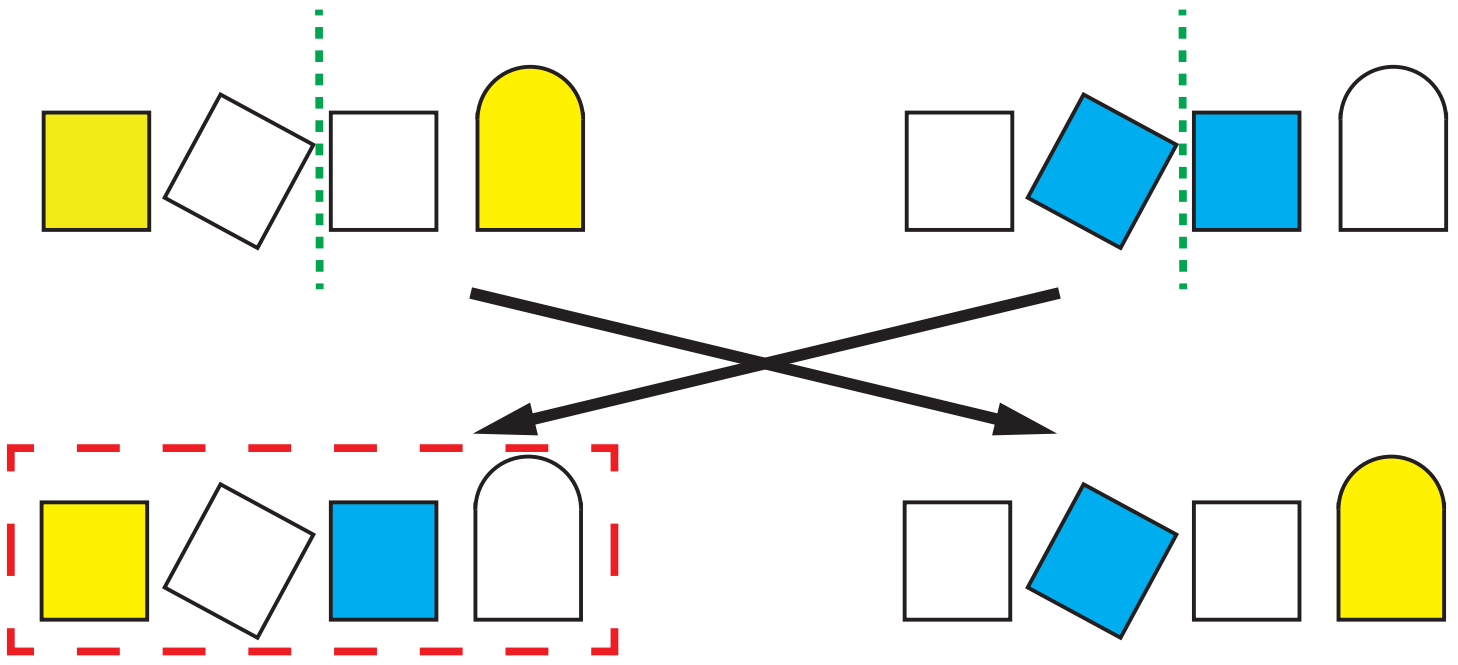

Supplementary Figure 2: Crossover operation in GA based subtomogram set refinement. Upper row, two parent solutions, where the colored shapes correspond to selected subtomograms, and white shapes correspond to unselected subtomograms. The dashed green line represents the crossover point. Lower row, two children solutions after applying crossover operation. Highlighted in dashed rectangle is a better solution where selected subtomograms contain same shape with same orientation.

Given currently aligned subtomograms, and a binary vector that indicates which subtomograms are selected, we can calculate a SFSC score defined in Equations 2 as in the Step 1 of the above process. Then the score can be directly used as fitness that determines how likely a candidate solution in $C^{i-1}$ can be selected for reproduction in Step 2a. Suppose $S = \{\tilde{\rho}(\mathbf{o}_1), \ldots, \tilde{\rho}(\mathbf{o}_{2n})\}$ are SFSC scores of the candidate solutions $\{\mathbf{o}_1, \ldots, \mathbf{o}_{2n}\}$ in the combined population $C^{i-1}$. Then the probability of selecting an individual candidate solution $\mathbf{o}_j$ is calculated as

$$P(\mathbf{o}_j) = \frac{\tilde{\rho}(\mathbf{o}_j) - s^{\min}}{\sum_k [\tilde{\rho}(\mathbf{o}_k) - s^{\min}]}, \ \forall 1 \leq j \leq 2n$$

where $s^{\min} := \min_l \tilde{\rho}(\mathbf{o}_l)$.

Remarks: In principle, the GA based subtomogram selection method can also be used as an alternative pattern generation method in the MPP framework (Supplementary Note 1.1.1). However, because the GA approach is significantly more time consuming compared to k-means clustering and sequential expansion (Supplementary Note 1.1.3.2) approaches, instead of integrating it into the MPP framework, we use it only for refining selected individual patterns predicted using MPP.

#### 1.1.4 Selection of disjoint high quality patterns

In contrast to a typical template-free subtomogram classification method, MPP is a constrained optimization method that improves a selection of distinct high quality patterns (in terms of SFSC scores, defined in Equation 2) from a pattern library, which contains not only the patterns from the current iteration but also patterns generated in any previous iteration. In such case the overall quality of selected patterns tend to increase with the advance of iterations until reaching convergence at which MPP can hardly improve the pattern quality.

In order to reduce the chance of selecting redundant patterns from the pattern library, we assume that one subtomogram generally can belong to no more than one selected pattern. In other words, we want the selected patterns to be disjoint in terms of their subtomogram set membership.

We propose an iterative greedy pattern selection process (as summarized in Algorithm 1). Such process keeps adding patterns into a collection $S$ from the pattern library $L$ based on several search criteria: 1) high quality patterns, 2) minimal overlap in subtomogram membership, and 3) maximal overall subtomogram coverage. This procedure ensures the selection of a disjoint set of patterns with minimal subtomogram overlap between them (i.e., subtomograms are not shared between patterns). First, all patterns in the library are ranked according to their pattern quality measure. Staring with the highest quality pattern, a pattern is added to the collection $S$ if it has the highest ranked quality among all patterns and with subtomogram member overlap smaller than a certain small threshold $t^{\text{overlap}} = 0.01$ with all the subtomograms of all already selected patterns part of the pattern collection $S$. To increase coverage, the process selects as many eligible patterns as possible, until no more eligible pattern can be found in the pattern library $L$.

**Algorithm 1** Select high quality patterns with minimal subtomogram set overlap

**Require:** A library $L$ of patterns $p_1, p_2, \ldots, p_{|L|}$ with corresponding subtomogram sets $C_{p_1}, C_{p_2}, \ldots, C_{p_{|L|}}$, and with corresponding SFSC scores in order: $\tilde{\rho}_{p_1} \geq \tilde{\rho}_{p_2} \geq \ldots \geq \tilde{\rho}_{p_{|L|}}$, a max overlap ratio $t^{\text{overlap}}$

$S \leftarrow \varnothing$
**for** $i \leftarrow 1$ to $|L|$ **do**
  $A \leftarrow \cup_{p \in S} C_p$
  **if** $|C_{p_i} \cap A| \leq t^{\text{overlap}} |C_{p_i}|$ **then**
    $S \leftarrow S \cup \{p_i\}$
**return** $S$

### 1.1.5 Align subtomogram averages into common frames

After selecting a disjoint set $S^{\text{sel}}$ of high quality patterns according to Supplementary Note 1.1.4, the corresponding pattern averages in $S^{\text{sel}}$ are aligned into common frames. This procedure helps the dimension reduction to focus more on the structural difference among the averages rather than the variance introduced due to orientation and location differences of patterns with similar structures. Such technique has been used in the align-and-classify frameworks [e.g. 5]. However, the alignment of all averages into a *single* common frame is not be appropriate for a visual proteomics setting, which contains structures of many different complexes of largely different shape and size. The alignment of two averages of largely different structures may be meaningless and can result in large displacements of one structure to outside the boundary of its subtomogram volume. To overcome this limit, we propose an alignment procedure that only aligns pairs of the structurally most similar averages. The procedure is summarized in Algorithm 2.

### 1.1.6 Identification of structurally redundant patterns

When the true set of structurally distinct patterns is unknown, an intuitive strategy is to over-partition the collection of subtomograms then identify and remove the patterns of redundant structures, so that such redundant patterns will never be selected or processed in future iterations.

**Algorithm 2** Alignment of cluster averages into common frames

**Require:** A set $S_0$ of patterns, with subtomogram sets $C_1, C_2, \ldots, C_{|S_0|}$, with corresponding subtomogram averages $a_1, a_2, \ldots, a_{|S_0|}$, and with corresponding SFSC scores in order: $\tilde{\rho}_1 \geq \tilde{\rho}_2 \geq \ldots \geq \tilde{\rho}_{|S_0|}$. Denote the alignment score and translation between $a_i$ and $a_j$ as $r_{i,j}$ and $\mathbf{a}_{i,j}$ respectively.

Select and order the alignment scores to the subtomogram average pairs $(i_1, j_1), (i_2, j_2), \ldots, (i_{n^{\text{pair}}}, j_{n^{\text{pair}}})$, such that $r_{i_1,j_1} \geq, r_{i_2,j_2}, \geq \ldots, \geq r_{i_{n^{\text{pair}}}, j_{n^{\text{pair}}}}$, $||\mathbf{a}_{i_p,j_p}||_2 \leq t^{\text{translation}}, \forall p$, and $i_p < j_p, \forall p$.
$S^{\text{fixed}} \leftarrow \varnothing$
$S^{\text{transformed}} \leftarrow \varnothing$
**for** $p \leftarrow 1$ to $|S_0|$ **do**
  **if** $i_p \notin S^{\text{transformed}}$ and $j_p \notin S^{\text{fixed}} \cup S^{\text{transformed}}$ **then**
    Apply a rigid transform of $a_{j_p}$ according to its alignment against $a_{i_p}$
    $S^{\text{fixed}} \leftarrow S^{\text{fixed}} \cup \{i_p\}$
    $S^{\text{transformed}} \leftarrow S^{\text{transformed}} \cup \{j_p\}$

In principle, one may intuitively select a single similarity cutoff between the averages to identify structurally redundant patterns. However, in a visual proteomics setting, for different pairs of macromolecular complexes, one has to consider different degrees of image and structural differences as a result of varying coverage (i.e. number of subtomograms that contain a complex) and varying sizes for different complexes. Two high resolution subtomogram averages (based on a large number of subtomograms) may show relatively subtle but true differences. On the other hand, two low resolution subtomogram averages with the same underlying structure may show relatively large but false differences due to fluctuations of noise or misalignments of the subtomograms. Therefore it would be difficult to properly choose a single similarity cutoff to define structural redundancy for all patterns. To overcome this limit, we determine structural redundancy by measuring the statistical discrimination ability of alignment scores through statistical hypothesis testing. This procedure allows more flexibility in detecting systematic differences between two groups of alignment scores generated by aligning a set of subtomograms against two pattern density averages .

We use a statistical test of consistency between set membership and alignment scores to automatically identify structurally redundant patterns. The design of our method is based on the following intuitions: Given a collection of candidate patterns selected in Supplementary Note 1.1.4, if a pattern has a distinct subtomogram average compared to other patterns and the average reflects the true underlying structure of the subtomograms of the pattern, we expect that the subtomograms of this pattern should *specifically* well align (in terms of alignment scores) to the average of the pattern, as compared to their alignment against averages of any other pattern. Otherwise, either the subtomogram average of this pattern does not reveal the underlying true structure, or it cannot be discriminated from the subtomogram average of some other patterns because both averages contain structures that are too similar to be discriminated by the alignment scores. We use such a statistical consistency between subtomogram set membership and alignment as a criterion to detect redundant patterns. This is useful for removing candidate patterns whose averages do not reflect the true underlying structure and candidate patterns of redundant structures (that are already considered by another pattern). With the removal of such patterns, the computational cost of MPP can be significantly reduced.

Formally, we define a pattern $p \in S$ as *structurally redundant* with respect to another pattern $p' \in S$, if it has the following properties: 1) $p$ has a lower SFSC score than $p'$, and 2) through an appropriate hypothesis testing, the alignment scores between the subtomograms of $p$ and the subtomogram average of $p$ is not significantly higher than the alignment scores between the subtomograms of $p$ and subtomogram average of $p'$. In such a case, the subtomogram average of $p'$ is likely to provide a better representation of the underlying structure in the subtomograms of $p$. Consequently, $p$ should be identified as redundant to $p'$ and be discarded from further processing.

More specifically, we propose a statistical test procedure to detect redundant patterns, which satisfies the above properties. Suppose at the current iteration, a collection of $S = \{p_1, \ldots, p_{|S|}\}$ of disjoint patterns have been selected according to Supplementary Note 1.1.4, and their corresponding subtomogram sets are denoted as $C_1, C_2, \ldots, C_{|S|}$. Their subtomogram averages are denoted as $a_1, a_2, \ldots, a_{|S|}$. Their corresponding SFSC scores are denoted as $\tilde{\rho}_1, \tilde{\rho}_2, \ldots, \tilde{\rho}_{|S|}$, and the patterns are ordered such that $\tilde{\rho}_1 \leq \tilde{\rho}_2 \leq \ldots \leq \tilde{\rho}_{|S|}$. Furthermore, let $r_{f,a_i}$ be the alignment score between a subtomogram $f$ and the average $a_i$. For any two patterns $p_i$ and $p_j$ with $i < j$, we compare the alignment scores $\mathbf{r}_{i,i} = (r_{f,a_i}, \forall f \in C_i)$ and $\mathbf{r}_{i,j} = (r_{f,a_j}, \forall f \in C_i)$ using Wilcoxon signed-rank test [85], which is a paired difference test. If $\mathbf{r}_{i,i}$ is not significantly higher than $\mathbf{r}_{i,j}$ (at a significance level of 0.01), then the subtomograms in $C_i$ do not align specifically well against $a_i$ compared with against $a_j$. In addition, since $\tilde{\rho}_i \leq \tilde{\rho}_j$, we identify $p_i$ as structurally redundant respect to $p_j$.

Remarks: Like any other statistical tests, our statistical test may fail when the number of subtomograms is small or when there are systematic bias in the alignment scores. It also depend on the discrimination ability of alignment scores.

#### 1.1.7 Reference guided automatic adaptive subtomogram masking and target complex region segmentation

Molecular crowding within cellular subvolumes has profound effects on macromolecular interactions [59, 76, 28] and makes visual proteomics scale analysis significantly more challenging. A subtomogram extracted from a tomogram of the crowded cell cytoplasm may not only contain the target complex of interest, but also some neighboring structures or structural fragments of other complexes. There are also cases that instances of purified complexes are crowded due to high concentration or clustering [e.g. 12]. The existence of neighboring structures and noise in the non-structural background regions inside subtomograms biases their alignments [101] and other processing steps of MPP.

To reduce the influence of noise at the background regions and the influence of neighboring structures on the subtomogram analysis, we propose an automatic method that uses a density map as a reference to segment the region occupied by the target complex, mask out regions occupied by neighboring structures, and partially mask out regions occupied by background noise. In the MPP framework the reference density map is a subtomogram average of a pattern selected based on the information on pairwise alignments of subtomograms against averages of the collection $S^{\text{remain}}$ of patterns (Supplementary Note 1.1.1). When using a reference as a seed, the method automatically identifies a region that includes the target complex with a margin that follows the shape of the target complex, and excludes the regions occupied by potential neighboring structures. This tool is an optional component of the MPP framework.

The basic idea of the procedure is illustrated in Supplementary Figure 3. Without loss of generality, we assume high image intensity of subtomograms corresponding to high electron density. Within a given MPP iteration (Supplementary Note 1.1.1), suppose the subtomogram $f$ is best aligned with pattern's average $a$ among all other averages of a collection of patterns $S^{\text{remain}}$. $f$ is smoothed using a Gaussian smoothing with $\sigma = 2$nm.

We first apply level set based segmentation on $a$ to identify structural region $R_a^{\text{structure}}$ . This is done according to Supplementary Note 1.2.1.

Once $a$ is segmented, we map the mask of the structured region $R_a^{\text{structure}}$ onto $f$ (Supplementary Figure 3b). We then calculate the mean intensity values of $f$ inside $R_a^{\text{structure}}$ and outside $R_a^{\text{structure}}$. Denote these two values as $c_1$ and $c_2$ respectively. We can then minimize the following model to obtain optimal level set as $\phi_f^*$ and structural region $R_f^{\text{structure}}$ in the similar way as done in Supplementary Note 1.2.1, except with fixed $c_1$ and $c_2$:

$$\phi_f^* = \arg\min_{\phi} \; \mu \int |\nabla H(\phi)| + \lambda \left[ \int |f - c_1|^2 H(\phi) + \int |f - c_2|^2 (1 - H(\phi)) \right]$$

We then separate the connected components of $R_f^{\text{structure}}$ into two groups: those that overlap with $R_a^{\text{structure}}$

and those that do not. The first group of connected components are defined as the structural regions of the target complex $R_f^{\text{target}}$. The second group of connected regions are defined as structural regions of neighboring structures $R_f^{\text{neighbor}}$. Then we perform Watershed segmentation [11, 95] on $\phi_f^*$ using $R_f^{\text{structure}}$ and $R_f^{\text{neighbor}}$ as initial seeds to partition the subtomogram into two regions $R_f^{\text{target_ext}}$ and $R_f^{\text{neighbor_ext}}$. The final target complex region mask is defined as $R_f^{\text{target_ext}} \cap \{\phi^* > t \max(\phi^*)\}$, where $t$ is a negative valued threshold parameter to control the amount of included margin. Such mask follows the shape of the target complex and excludes neighboring structures (Supplementary Figure 3c).

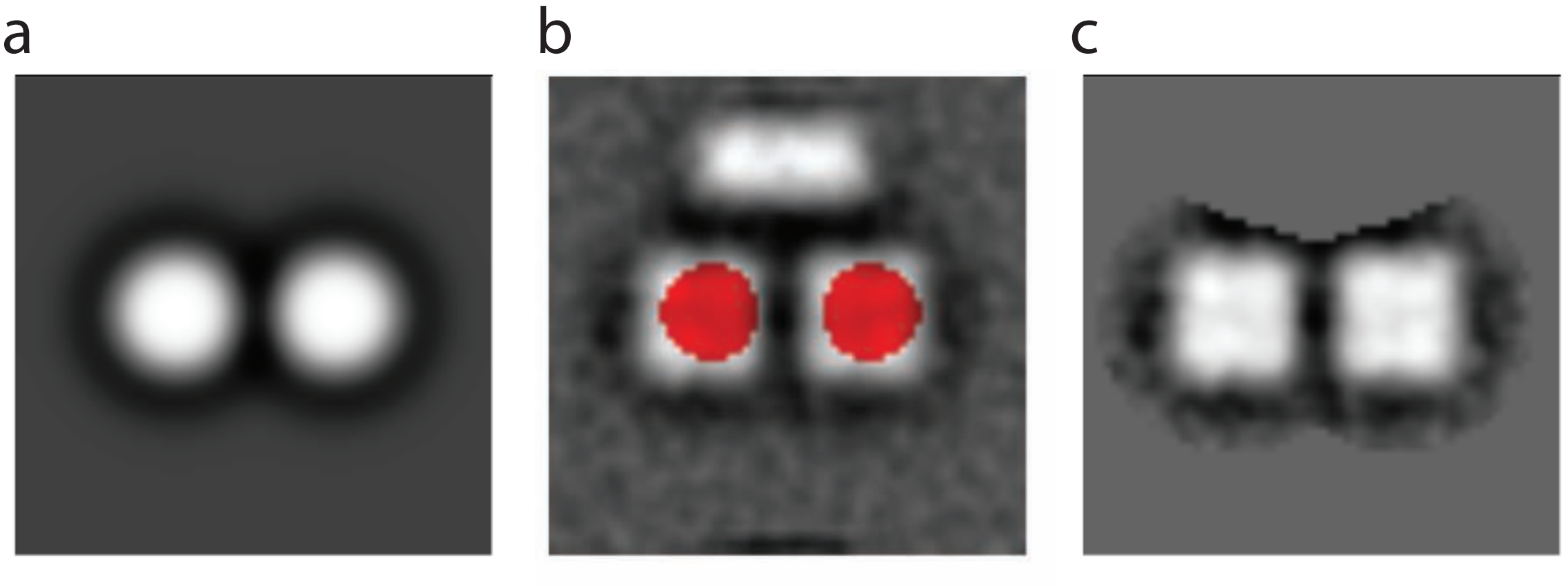

Supplementary Figure 3: Basic idea of reference guided segmentation is illustrated using a toy example. (a) A reference density map $a$. (b) A subtomogram $f$ that is roughly aligned against $a$. It contains a particle represented by two disjoint cubes. And a rectangular neighboring structure. The red region is a seed corresponding to $R_a^{\text{structure}}$ calculated from segmenting $a$. (c) Final masked subtomogram.

Remarks:

- The existence of neighboring structures besides the target complex in a subtomogram $f$ affects its alignment against a reference $a$ [101]. However, $a$ is only used as an initial seed. Therefore, even if the alignment is not accurate or the $a$ do not have the same structure as the underlying structure of $f$, as long as after alignment $R_a^{\text{structure}}$ overlaps with the true target complex region of $f$, and does not overlap with the neighboring structure region of $f$, we may still expect a successful segmentation.

- As illustrated in Supplementary Figure 3, even if target complex regions of $f$ are apparently disconnected, as long as the target complex regions of $f$ have overlap with $R_a^{\text{structure}}$, the disconnected subunits will still be included in the final segmentation.

- The reason of applying the watershed segmentation on $\phi_f^*$ instead of $f$ is because $\phi_f^*$ is derived from the distance transform [44], which represents the signed distances of voxel locations to the structural regions. $\phi_f^*$ is usually much more smooth than the noisy $f$. In addition, $\phi_f^*$ is a signed distance function that monotonically decreases when the distance to the segment increases. By contrast, due to the suppression of low frequency components in the CTF during the imaging process, $f$ has both above and below background intensity around the surface regions of structures. Therefore the segmented boundary from the watershed segmentation on $\phi_f^*$ would be much more regular than those from watershed segmentation directly on $f$.

- Due to its complexity, the segmentation of the target complex region is a very challenging problem when applied on a proteome scale. Many factors may lead to the failure of our reference guided segmentation approach. For example, the high degree of distortions in a subtomogram or high degree of misalignments of subtomograms against the reference may lead to under or over segmentation. If a subtomogram is highly crowded, some neighboring structures may appear to be connected with the target complex in the subtomogram, which makes the segmentation unable to exclude the neighboring structure region.

- In order to avoid false segmentation of a subtomogram average when it is very noisy, we assume that $R_a^{\text{structure}}$ has less overlap with the boundary of the subvolume than the non-structural region $(R_a^{\text{structure}})^C$ does, and use this assumption to discard bad segmentations.

#### 1.1.8 High frequency noise reduction of subtomogram averages using adaptive Gaussian smoothing

Sometimes, repeated iterations of alignment and averaging give a structure containing high resolution features resulting from the alignment of noise against itself in a reinforcing manner [13]. Such phenomena is called *over-alignment* [13]. In such case, it is beneficial to have an optional step to reduce high frequency noise.

Gaussian smoothing is a commonly used noise reduction technique. A number of studies in image analysis field show that within the class of linear transformations, Gaussian kernel minimizes the chance of creating new structures in the transformation from a finer to a coarser scale [52, 51, 53, 54, 56, 55, 78, 87]. We apply Gaussian smoothing to an average to reduce influence of noise, which is equivalent to applying a Gaussian envelope function in Fourier space. Such envelope function has the form of

$$f_{a,c}(x) = a \exp\left(-\frac{x^2}{2c^2}\right)$$

Since our procedure includes estimation of SSNR and FSC, the parameters $a$ and $c$ can be adaptively determined from the estimated FSC through least-squares fitting using the Levenberg-Marquardt algorithm [70].

## 1.2 Level set based pose normalization for pre-filtering

MPP (Supplementary Note 1.1.1) is suitable for processing thousands to tens of thousands of subtomograms with affordable computation cost. However, with the advance of automation of ECT imaging techniques [e.g. 65], nowadays it is not difficult to acquire a substantially larger amount (for example, more than a million) of subtomograms within a day. Using MPP alone is not computationally feasible for processing such a large amount of subtomograms. Therefore, an efficient coarse filtering of the subtomograms is very useful to reduce the whole collection of subtomograms to substantially smaller subsets containing structures of relatively similar sizes and shapes. Then these subsets can be further independently processed using MPP. In this paper, we perform such filtering through normalization of translation and rotation of subtomograms followed by clustering.

Intuitively, the normalization of the translation of the particle inside a subtomogram can be done by calculating a key point with respect to the particle, which is invariant to the rotation and translation of the particle. A typical example of such a keypoint is the center of mass. However, because the suppression of zero frequency signal in the ECT imaging process, the mean intensity value of a subtomogram is often close to the background intensity value [101]. Therefore it is hard to directly use all image intensities of a subtomogram to accurately estimate a center of mass of the particle. Instead, we use binary segmentation to obtain a coarse shape of the particle, and calculate the center of mass of this shape. Level set based segmentation [17] is a powerful, flexible method that can successfully segment many types of images, including some that would be difficult or impossible to segment with classical thresholding or gradient-based methods [23]. Through such segmentation, a coarse shape of the particle can be represented by the zero level region of a level set. The normalization of translation can then be calculated on the center of mass of the positive part of the level set instead of on the original image intensities. Given such center of mass, we can further estimate the general orientation (without taking into account missing wedge effect) of the particle by calculating the principal directions by applying PCA to the coarse shape (Supplementary Figure 4).

Such pose normalization procedure can be independently and efficiently applied to individual subtomograms. With the coarse alignment from pose normalization, it is possible to separate particles with very distinct sizes and distinct elongated shapes through simple and efficient clustering techniques like k-means clustering and generate an average representing of such general shapes. Then averages of subtomograms can be inspected and the corresponding subtomogram sets can be selected for further more focused analysis. Such procedure is highly scalable and can be easily parallelized. It can usually process tens of thousands of subtomograms on a single computer within one day.

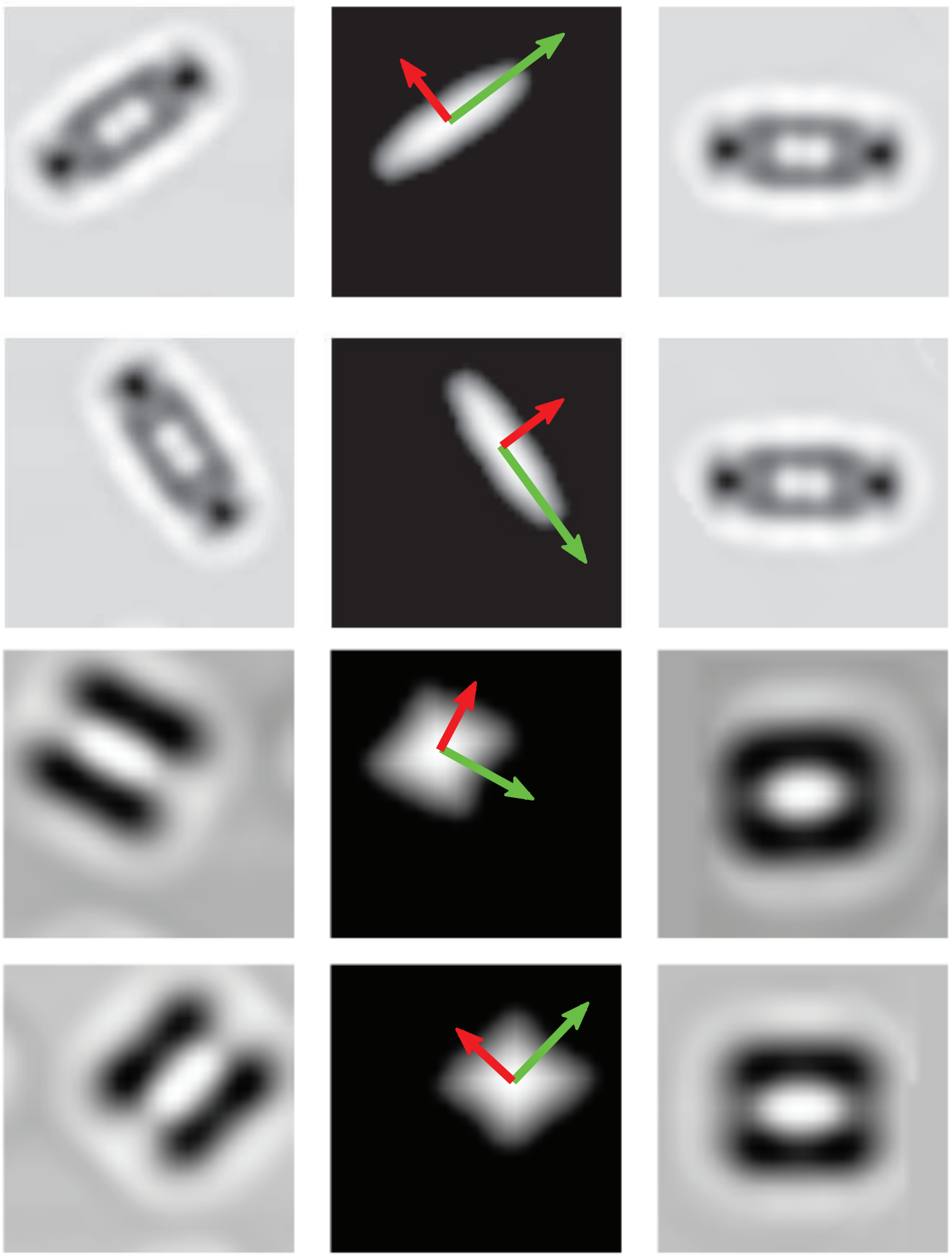

Supplementary Figure 4: Basic idea of level set based pose normalization is illustrated using four toy examples in four rows. Left column, center slice of four simulated subtomograms containing two Proteosome and two GroEL complexes with different orientation and location. Middle column, the density map is the positive part of the approximation level sets, and the vectors are inferred pose. Right column, pose normalized subtomograms.

### 1.2.1 Level set approximation and structural region segmentation

We formulate the identification of structural regions as a binary region based segmentation problem that minimizes the Chan-Vese model [17], which is a popular level set based segmentation model. The model can be formulated as follows.

$$\underset{c_1,c_2,\phi}{\arg\min}\ \mu\int|\nabla H(\phi)|+\lambda\left[\int|f-c_1|^2H(\phi)+\int|f-c_2|^2(1-H(\phi))\right] \tag{5}$$

In Equation 5, $f:\mathbb{R}^3\to\mathbb{R}$ is the intensity of the subtomogram to be segmented. $\phi:\mathbb{R}^3\to\mathbb{R}$ is a level set function that simultaneously defines a boundary contour and a segmentation of an image. The boundary contour is taken to be the zero level set $\{\phi=0\}$, and the segmentation is given by the two regions $\{\phi<0\}$ and $\{\phi\geq 0\}$. $H$ is the Heaviside step function $H(x)=\begin{cases}0, & x<0,\\ 1, & x\geq 0,\end{cases}$. $c_1$ and $c_2$ are the mean intensities inside the two regions.

The first term in Equation 5 measures the total area of the segment boundary. The minimization of the first term encourages the resulting segment boundary to be smooth. The second term measures the difference between image intensity and the mean intensity of the corresponding segments. The minimization of the second term encourages the uniformity of the intensities inside the two regions.

Such optimization problem can be elegantly solved by evolving the level set function $\phi$ through variational calculus [17]. In practice, we use [107] as an implementation of the algorithm, where $\phi$ is implemented using distance transform [44]. For simplification, we choose $\mu=1$, and $\lambda=\frac{1}{\mathrm{Var}(f)}$, where $\mathrm{Var}(f)$ is variance of $f$.

Let $\phi^*$ be the optimal level set. Suppose the region $R^{\text{structure}}=\{\phi^*>0\}$ corresponds to the high electron density in the subtomogram, then $R^{\text{structure}}$ is used to define the structural region inside the subtomogram.

Remark: In order to reduce the influence of noise, we usually apply a Gaussian smoothing with $\sigma=2$nm to a subtomogram before segmentation.

### 1.2.2 Pose normalization

The pose normalization is performed according to the positive part of $\phi^*$. Let $\phi_1^*(\mathbf{x})=\mathbf{1}_{\phi^*(\mathbf{x})\geq 0}\ \phi^*(\mathbf{x})$, where $\mathbf{1}$ is the indicator function. The pose normalization consist of following steps: First, we calculate a center of mass $\mathbf{c}_{\phi_1^*}$ of $\phi_1^*$.

$$\mathbf{c}_{\phi_1^*}=\frac{\int_{\mathbf{x}}\phi_1^*(\mathbf{x})\mathbf{x}}{\int_{\mathbf{x}}\phi_1^*(\mathbf{x})}$$

Then, we calculate

$$\mathbf{W}=\int_{\mathbf{x}}[\phi_1^*(\mathbf{x})]^2(\mathbf{x}-\mathbf{c}_{\phi_1^*})(\mathbf{x}-\mathbf{c}_{\phi_1^*})^\top \tag{6}$$

Then we calculate the eigen decomposition $\mathbf{W}=\mathbf{Q}\mathbf{\Lambda}\mathbf{Q}^\top$ of $\mathbf{W}$, where $\mathbf{Q}$ is an orthogonal matrix consists of eigenvectors, and the magnitude of eigenvalues in the diagonal matrix $\mathbf{\Lambda}$ are ordered in descending order. Finally, the pose normalization is performed by first translating the subtomogram (masked with $R^{\text{target_ext}}$) from $\mathbf{c}_{\phi_1^*}$ to the center of the subtomogram, then rotating the subtomogram using $\mathbf{Q}$ as rotation matrix.

Remarks:

- The coarse filtering of subtomograms may also be performed through rotation invariant features [100, 102, 29, 20, 18] combined with clustering. However, to extract structure information from such filtering is not straightforward because rotation invariant features do not provide alignment information. By contrast, for pose normalized subtomograms, coarse representative shapes can be directly obtained from cluster centers or subtomogram averages, which is very useful for manual inspection of these clusters.

- This method may not work when the SNR or contrast is very low. In addition, how to incorporate missing wedge to achieve a better pose estimation is an open problem.

# 2 Results

## 2.1 Tests of MPP on simulation data

### 2.1.1 Simulation of tomographic imaging process

For a reliable assessment of the method, simulated tomograms and subtomograms are generated by simulating the actual tomographic image reconstruction process, allowing the inclusion of noise, tomographic distortions due to missing wedge, and electron optical factors such as Contrast Transfer Function (CTF) and Modulation Transfer Function (MTF). We follow a previously applied methodology for realistic simulation of the tomographic image formation processs [8, 32, 68, 102].

The electron optical density of a macromolecule is proportional to its electrostatic potential and the density map can be calculated from the atomic structure by applying a low pass filter at a given resolution. An initial density map is then used as a sample for simulating electron micrograph images at different tilt angles. In ECT the sample is tilted in small increments around a single-axis. At each tilt angle a simulated micrograph is generated from the sample. In the real imaging process, the tilt angle range is limited. Therefore our data contain a wedge-shaped region in Fourier space for which no structure factors have been measured (i.e., the missing wedge effect). The missing wedge effect leads to distortions of the density maps. These distortions depend on the structure of the object and its orientation with respect to the direction of the tilt-axis. To generate realistic micrographs, noise is added to the images according to a given SNR level, defined as the ratio between the variances of the signal and noise [32]. Moreover, the CTF and MTF models distortions from interactions between electrons and the specimen and distortions due to the image detector [34, 68] in a linear approximation. Therefore, the resulting image is convoluted with a CTF. Any negative contrast values beyond the first zero of the CTF are eliminated. We also consider the MTF of a typical detector used in tomography, and convolute the density map with the corresponding MTF. Typical acquisition parameters that were also used during actual experimental measurements were used: voxel size = 1nm, the spherical aberration = 2.2mm, the defocus value = -15$\mu$m, the voltage = 300kV, the MTF corresponded to a realistic electron detector [60], defined as $\text{sinc}(\pi\omega/2)$ where $\omega$ is the fraction of the Nyquist frequency. Finally, we use a backprojection algorithm [68] to generate a tomogram or a subtomogram from the individual 2D micrographs that were generated at the various tilt angles [8, 102].

### 2.1.2 Validation procedure

To measure the performance of MPP, we calculate several quantities for comparing the prediction with ground truth. The first quantity is the membership consistency in terms of the amount of subtomogram membership overlap between a predicted pattern and the true set of a complex. Such membership consistency is represented as a contingency table. We order the columns and rows in the contingency table by identifying best matching using the Hungarian algorithm [46]. In an ideal case, when properly ordered, such a table would have non-zero entries in the diagonal cells, and zeros elsewhere.

Second, we also calculate the False Positive (FP) and False Negatives (FN) to measure the amount of instances (i.e. subtomograms) that MPP cannot correctly identify. Suppose, by checking the diagonal entry of a rearranged contingency table, the best matching between complexes and patterns is determined. Suppose a complex $c$ matches a pattern $p$. The FP of $p$ is the number of instances of pattern $p$ that do not belong to $c$, although they are predicted as instances of $c$ (because they are in $p$ and $p$ matches $c$). Given FP, we further calculate the False Discovery Rate (FDR) as FP divided by the total number of instances of $p$. The FDR indicates the level of impurity of $p$. The FN of $c$ is the number of true instances of $c$ that are not included into $p$. Given FN, we also calculate the Miss Rate or False Negative Rate (FNR) as FN divided by the total number of instances in $c$. Note that if $p$ correctly predicted the structure of $c$, in principle the missed instances (which are counted as false negatives) can be later detected through a template search.

Third, we also calculate the structural consistency between the average density map of a pattern and the true density map of the target complex. The consistency is measured in terms of FSC with 0.5 cutoff, which reflects the minimum scale that the predicted and true structures are consistently determined by the cutoff.

### 2.1.3 Tests on individually simulated subtomograms

In this section, we assess MPP with simulated subtomograms imitating macromolecular complexes extracted under very low crowding conditions, as they would be observed from relatively thin samples of purified complexes or cell extracts with very low concentration.

We randomly selected a collection of PDB structures of 22 macromolecular complexes (Supplementary Data 1a) of distinct shapes and sizes. The structures were converted into density maps using the pdb2vol program in the situs package [97] at 1nm voxel spacing and band pass filtered at 4nm. The density maps served as input for realistically simulating the cryo electron imaging process an SNR of 0.005 and tilt angle range $\pm 60°$ (see Supplementary Note 2.1.1 for details). For each complex 1000 subtomograms were generated, each containing a randomly rotated and translated complex. We then selected a random copy number (uniformly sampled from 1 to 1000) of simulated subtomograms for each complex. In total we collected 11,230 subtomograms as input data set for MPP (Supplementary Data 1a).

The MPP analysis is run in two stages, which differ in the generation of patterns at each iteration. In stage 1, the cluster number for the k-means clustering is set to $k^{\text{k_means_fix}} = 40$. A slight variation of $k^{\text{k_means_fix}}$ does not change the outcome of our analysis. The distribution of SFSC scores of patterns, and types of patterns over iterations are shown in Supplementary Figure 5. In the early iterations (iterations 0 to 14), the selected patterns have relative low quality, and there are large amounts of redundant patterns being detected. During the MPP iterations, the overall quality of the selected patterns increase. After iterations 15 to 19 in the stage 1 the improvement of quality of the selected patterns has reached its limit, and no new non-redundant patterns are selected. After iteration 20 we switch to stage 2, and because the subtomogram set sizes become less restricted due to adaptive k-means and sequential expansion modes, we start to see a further improvement of the quality of generated and selected patterns. The selected patterns continue to be improved during iterations 20 to 26, and then converge after iteration 27. Stage 2 is then terminated at iteration 32.

Computation time: When running MPP on 300 parallel workers with 40 selected patterns, one iteration took about 7 hours. Pairwise alignment between subtomograms and selected patterns is the most time consuming step, which took about 6 hours.

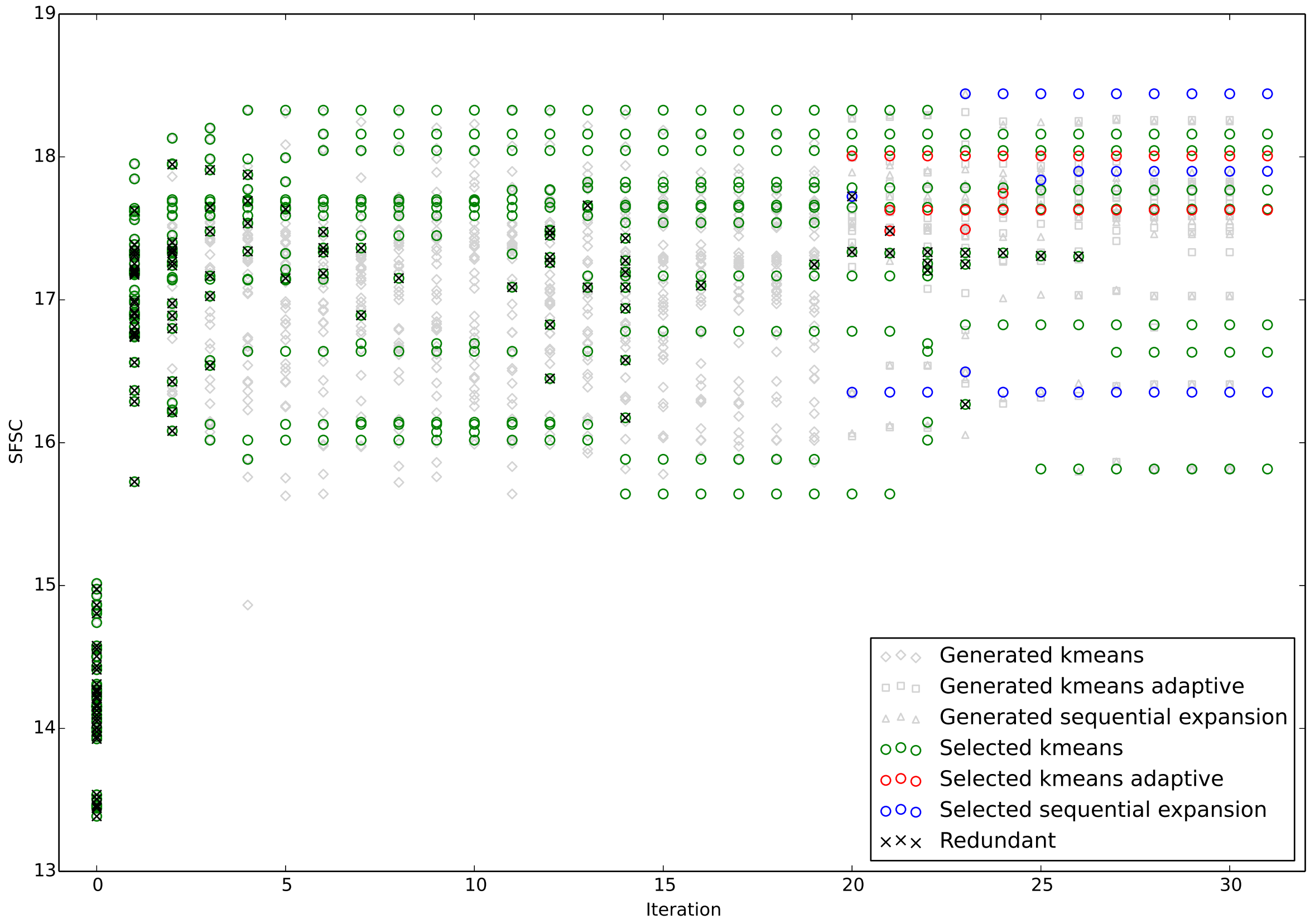

Supplementary Figure 5: The SFSC scores of different types of patterns over iterations. The light gray shapes correspond to the generated patterns being added into the pattern library. The colored circles correspond to patterns selected from the pattern library. The colored circles with cross correspond to selected but redundant patterns to be removed from the pattern library and omitted from further processing.

#### 2.1.3.1 Additional tests

**2.1.3.1.1 Different initial orientations** To test the reproducibility, we have repeated the test of MPP using same data and parameters as in Supplementary Note 2.1.3, but with different initial orientations. The result is very similar to the original test (Supplementary Figure 6 and Supplementary Data 2). Except, there are 4 redundant patterns (ID 3, 9, 10 and 12) not identified and removed.

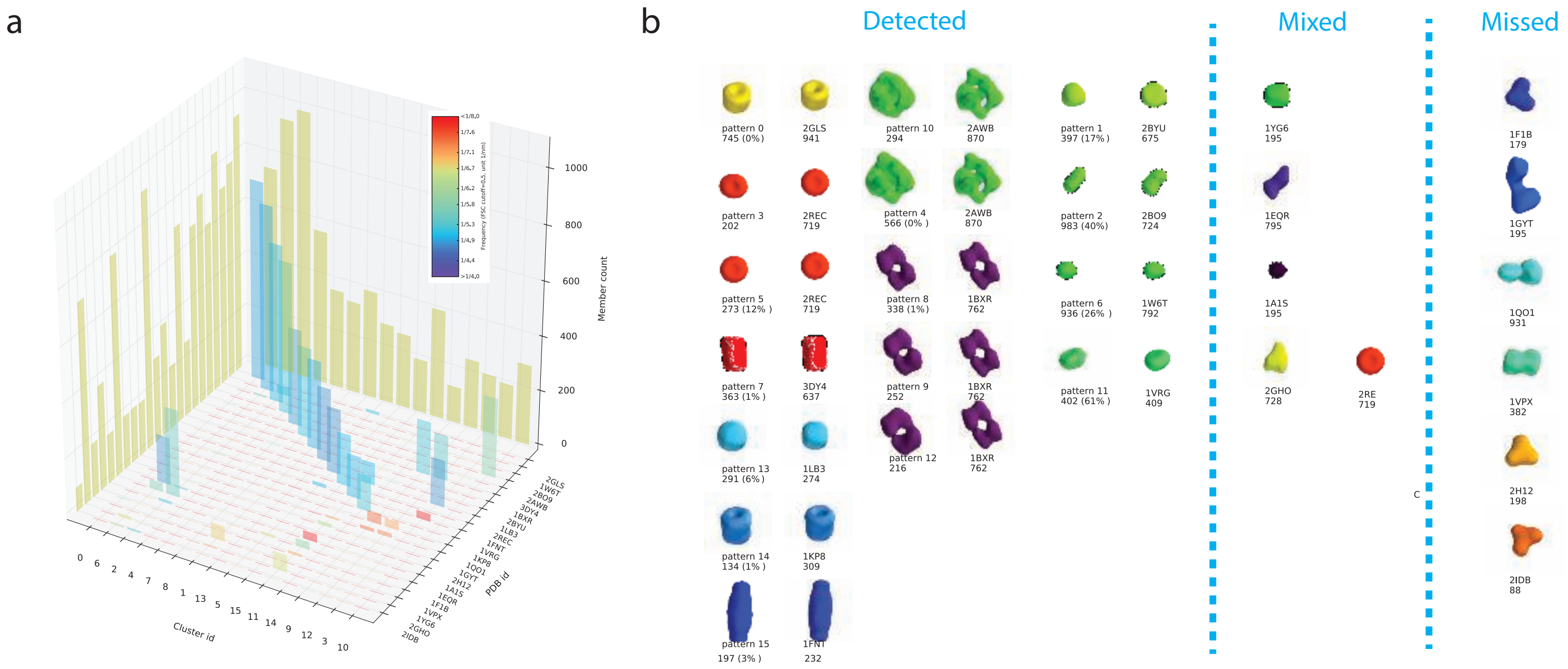

Supplementary Figure 6: Comparison of MPP pattern mining results with ground truth complexes for individually simulated subtomograms. (a) Results of individually simulated subtomograms, column plot representation of the contingency table (Supplementary Data 2) of the subtomogram membership overlap between true and inferred patterns. The height of each column corresponds to the number of subtomograms of the ground truth complex and the predicted patterns. The colors of the columns indicate the structural consistancy between the ground truth and the corresponding pattern averages, quantified as FSC with cutoff 0.5 (Supplementary Section2.1.2). (b) The isosurfaces of predicted patterns compared to ground truth structures. The ground truth structures are indicated by their PDB ID code, and the number of instances. Also shown are the isosurface representations of the predicted patterns with the number of instances and the false discovery rate (FDR) in parentheses. The FDR shows the fraction of wrongly assigned subtomograms in the pattern.

**2.1.3.1.2 Different abundance distribution** Furthermore, we have tested MPP using subtomograms of macromolecular complexes with a different abundance distribution. The result (Supplementary Figure 7 and Supplementary Data 3) shows that MPP can discover 4 patterns (ID 0, 2, 8, and 10) with one-to-one correspondence to ground truth in terms of both membership consistency and structural consistency. One complex (PDB ID 1BXR) has been predicted in two patterns (ID 3 and 7) with correct structure and without mixing with other complexes. There is one pattern (ID 4) that has correctly predicted one-to-one subtomogram membership of a complex (PDB ID: 1LB3) but wrongly predicted structure. There are two patterns (ID 1 and 6) whose subtomograms are a mixture of 4 complexes (PDB ID: 2BO9, 1W6T, 1A1S, and 1EQR), which has very similar structure (Supplementary Figure 10). There is one pattern (ID 5) whose subtomogram is a mixture of three complexes (PDB ID: 2BYU, 1QO1, and 1YG6), among which one complex (PDB ID 1QO1) has wrongly predicted structure. There are 7 missed complexes, 6 of them are of relatively low abundance (2AWB:320, 1F1B:719, 1FNT:110, 2GHO:46, 2H12:172, 2IDB:310, 2REC:183), again indicating that low abundance is a major factor for missing from mining.

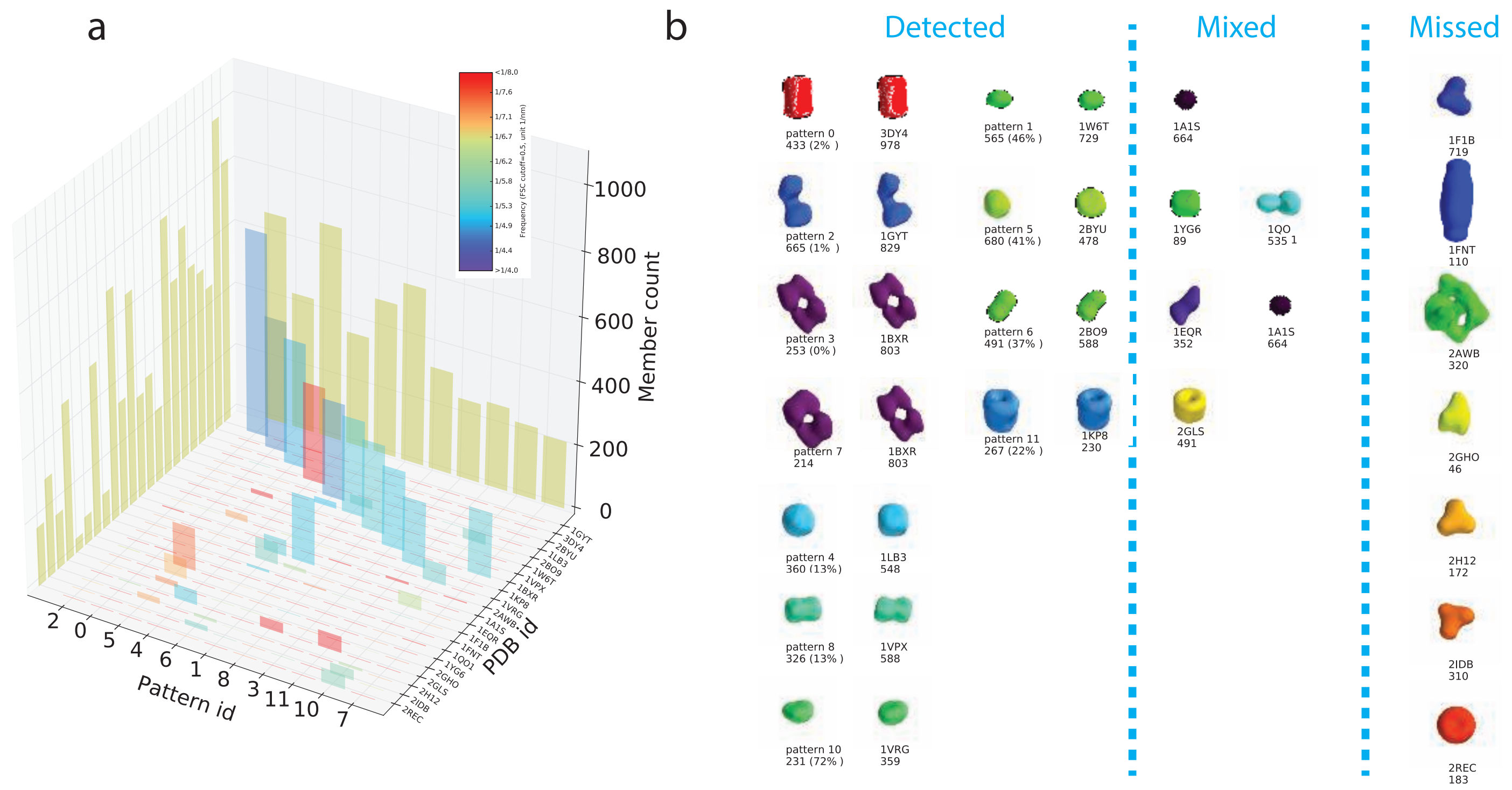

Supplementary Figure 7: Comparison of MPP pattern mining results with ground truth complexes for individually simulated subtomograms. (a) Column plot representation of the contingency table (Supplementary Data 3) of the subtomogram membership overlap between true and inferred patterns. The height of each column corresponds to the number of subtomograms of the ground truth complex and the predicted patterns. The colors of the columns indicate the structural consistency between the ground truth and the corresponding pattern averages, quantified as FSC with cutoff 0.5 (Supplementary Section2.1.2). (b) The isosurfaces of predicted patterns compared to ground truth structures. The ground truth structures are indicated by their PDB ID code, and the number of instances. Also shown are the isosurface representations of the predicted patterns with the number of instances and the false discovery rate (FDR) in parentheses. The FDR shows the fraction of wrongly assigned subtomograms in the pattern.

### 2.1.4 Tests on subtomograms extracted from simulated whole tomogram

#### 2.1.4.1 Simulation of tomograms containing a crowded mixture of macromolecular complexes

A density map is generated for each complex (the collection of 22 complexes used in Supplementary Note 2.1.3) at 1nm voxel spacing and band pass filter the map at 4nm. We apply level set based segmentation (Supplementary Note 1.2.1) on the density map of each complex. For each segment, we calculate a minimum bounding sphere, which is the smallest sphere that encloses the segment. We randomly place non-overlapping bounding spheres of 10,000 instances of the 22 complexes (with various abundance per type) into a volume $V$ of size $600 \times 600 \times 200\text{nm}^3$. Overlap between bounding spheres is prevented by applying molecular dynamics simulations in combination with an excluded volume constraints for all bounding spheres [81, 73]. Finally we embed the density maps of each randomly oriented complex into the $V$ according to locations of their corresponding bounding spheres. The combined large density map of all complexes had a crowding level (in terms of volume occupancy) of 15.2%, which is within the volume occupancy range (from 5% to 44%) that have been observed in cell cytoplasm [27, 28, 94, 38]. The density map of the crowded protein complexes is used to simulate a tomogram at SNR 50 and tilt angle range $\pm 60°$ according to Supplementary Note 2.1.1 (Supplementary Figrue 8).

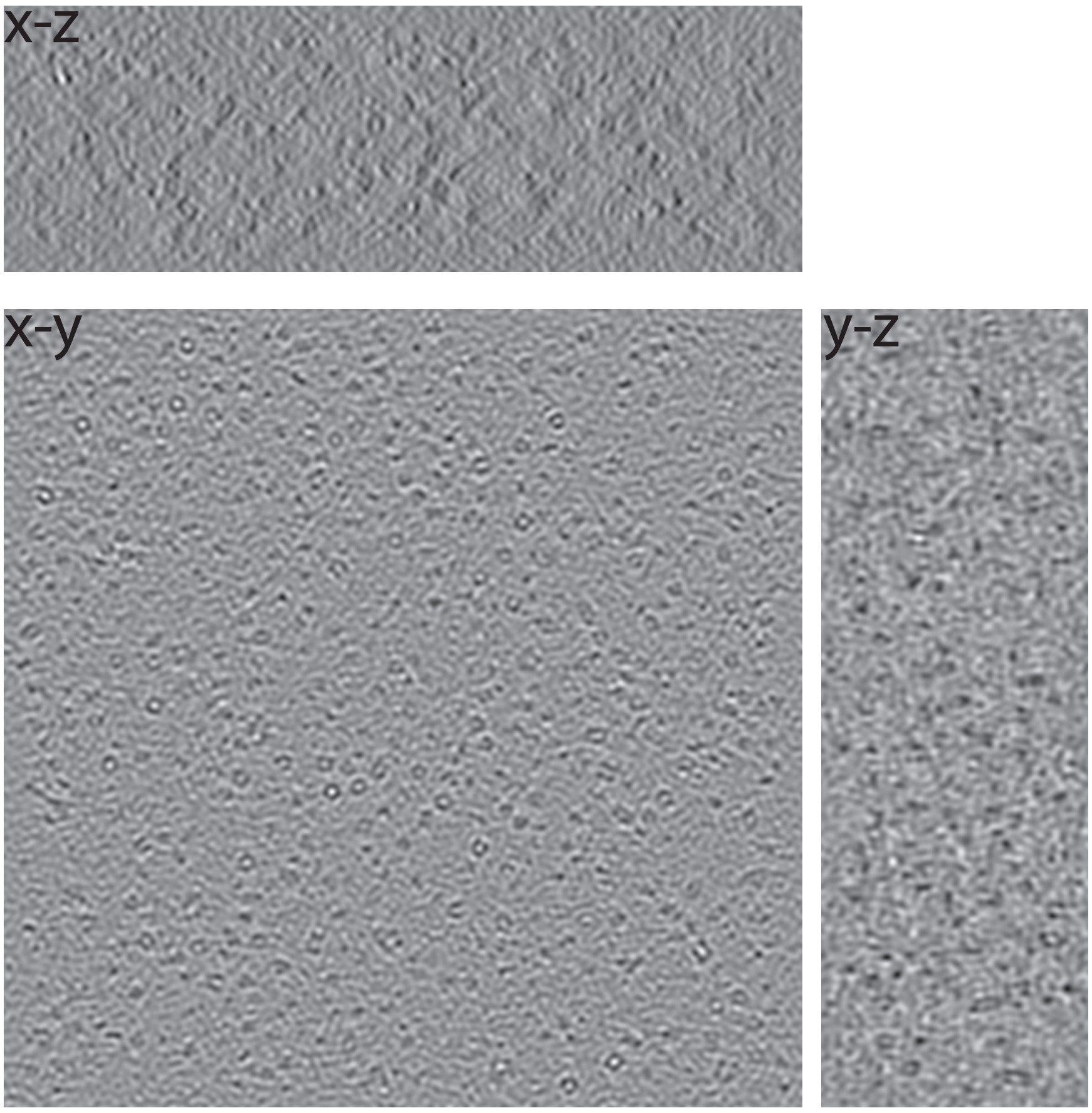

Supplementary Figure 8: Three slices of the simulated crowded tomogram.

**2.1.4.2 Particle picking and subtomogram extraction** The subtomogram extraction is done through template-free particle picking. The particle picking is based on Difference of Gaussian (DoG) filtering [96]. We filter a tomogram $v_1$ using a DoG function of $\sigma = 7$nm and K-factor = 1.1, resulting in a filtered tomogram $f_1$. We search for the collection $p_1$ of local maxima peaks of $f_1$. Often, there are false positive peaks, i.e. those peaks that do not correspond to macromolecular complex instances but rather noisy fluctuations in the non-structural regions. To reduce such false peaks in $p_1$, we randomly sample voxels to form another volume $v_0$ of size $400 \times 400 \times 200\text{nm}^3$. Then we apply the same DoG filtering to obtain a filtered map $f_0$. Then we perform local maxima search to obtain a collection $p_0$ of background peaks. Finally we selected a total of 4,901 peaks in $p_1$ whose values are larger than 5 times of the standard deviation plus mean of the values of $p_0$. In order to evaluate the performance, we identify true class labels of these peaks through the one-to-one correspondence between peak locations and the minimal bounding spheres. As a result, 4,618 peaks are assigned true class labels (Supplementary Data 4a). Due to the size preference of DoG particle picking, when setting $\sigma = 7$nm, instances of large complexes are more likely to be picked out than instances of small complexes. Centered at each of the 4,901 peaks, we cut out a subtomogram of size $30^3$ voxels. These subtomograms are used as an input of MPP.

Remarks: In this paper, for simplicity, we use DoG with a single fixed $\sigma$ for particle picking. DoG particle picking has size preference of picked particles. In practice, in order to detect patterns of very different sizes, one may systematically perform DoG particle picking using multiple $\sigma$, followed by pattern mining. In addition, other types of template-free particle picking methods may be used instead of using DoG particle picking.

**2.1.4.3 Test of MPP on highly crowded protein complex mixture** After extracting the 4,901 subtomograms, we apply the MPP procedure (Supplementary Note 1.1.1) to the extracted subtomograms using similar settings as in Supplementary Note 2.1.3. During the MPP iterations, we appled our reference guided segmentation (Supplementary Note 1.1.7) to reduce the influence of crowdedness. Supplementary Data 4b summarizes the resulting patterns.

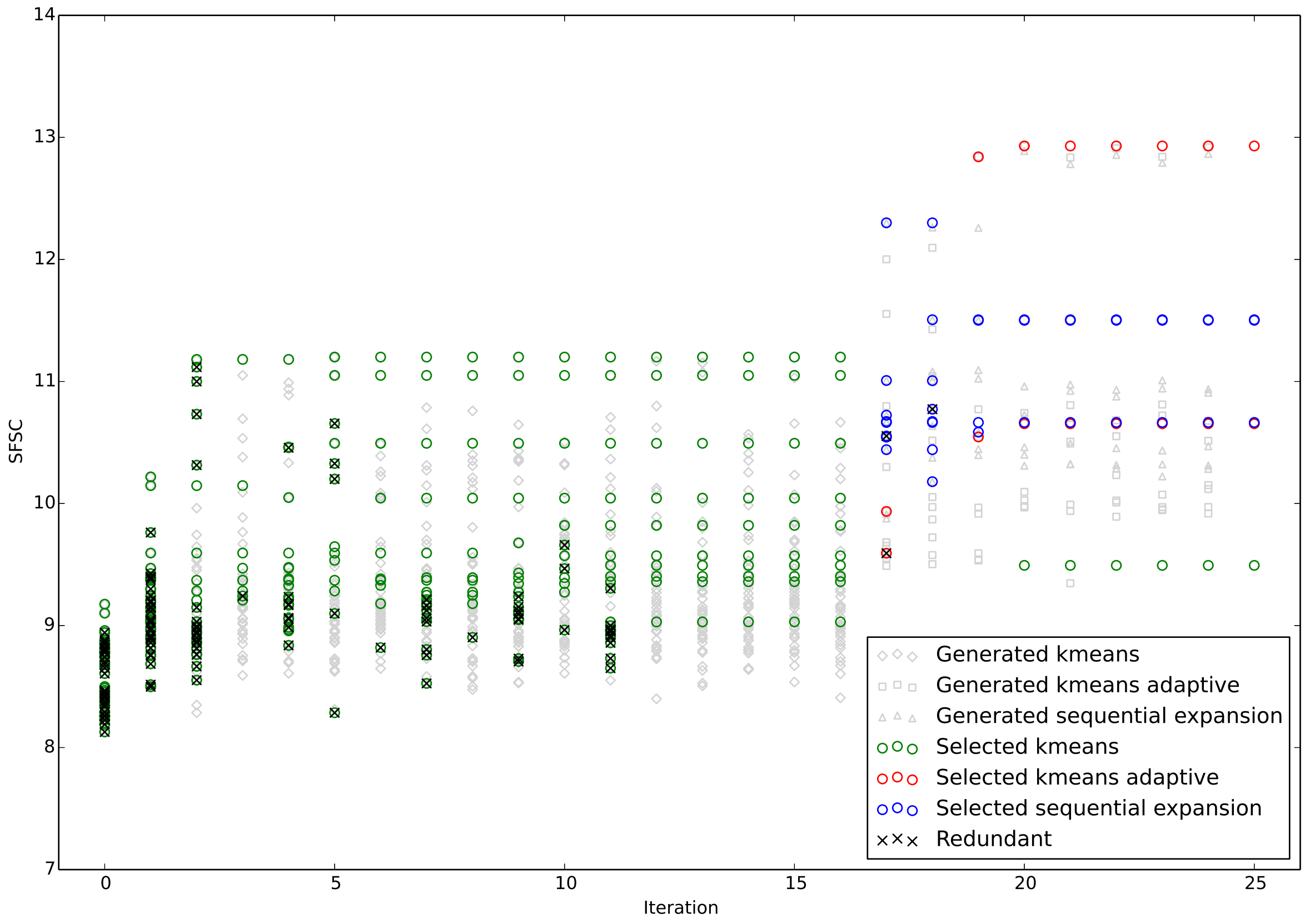

Supplementary Figure 9: The SFSC scores of different types of patterns over iterations. The light gray shapes correspond to the generated patterns being added into the pattern library. The colored circles correspond to patterns selected from the pattern library. The colored circles with cross correspond to selected but redundant patterns to be removed from the pattern library and omitted from further processing.

**2.1.4.3.1 Estimation of SNR level of subtomograms** We sampled 10,000 pairs of aligned subtomograms of pattern 4, which are dominated by GroEL complex. For each pair of subtomograms, we calculate Pearson correlation of their image intensity, then estimate corresponding SNR according to [35]. Such procedure gives an SNR estimate of $0.29 \pm 0.13$ over all subtomograms pairs, which is of similar range to the one estimated form the *A. longum* cellular tomogram (Supplementary Note 2.2.1.5).

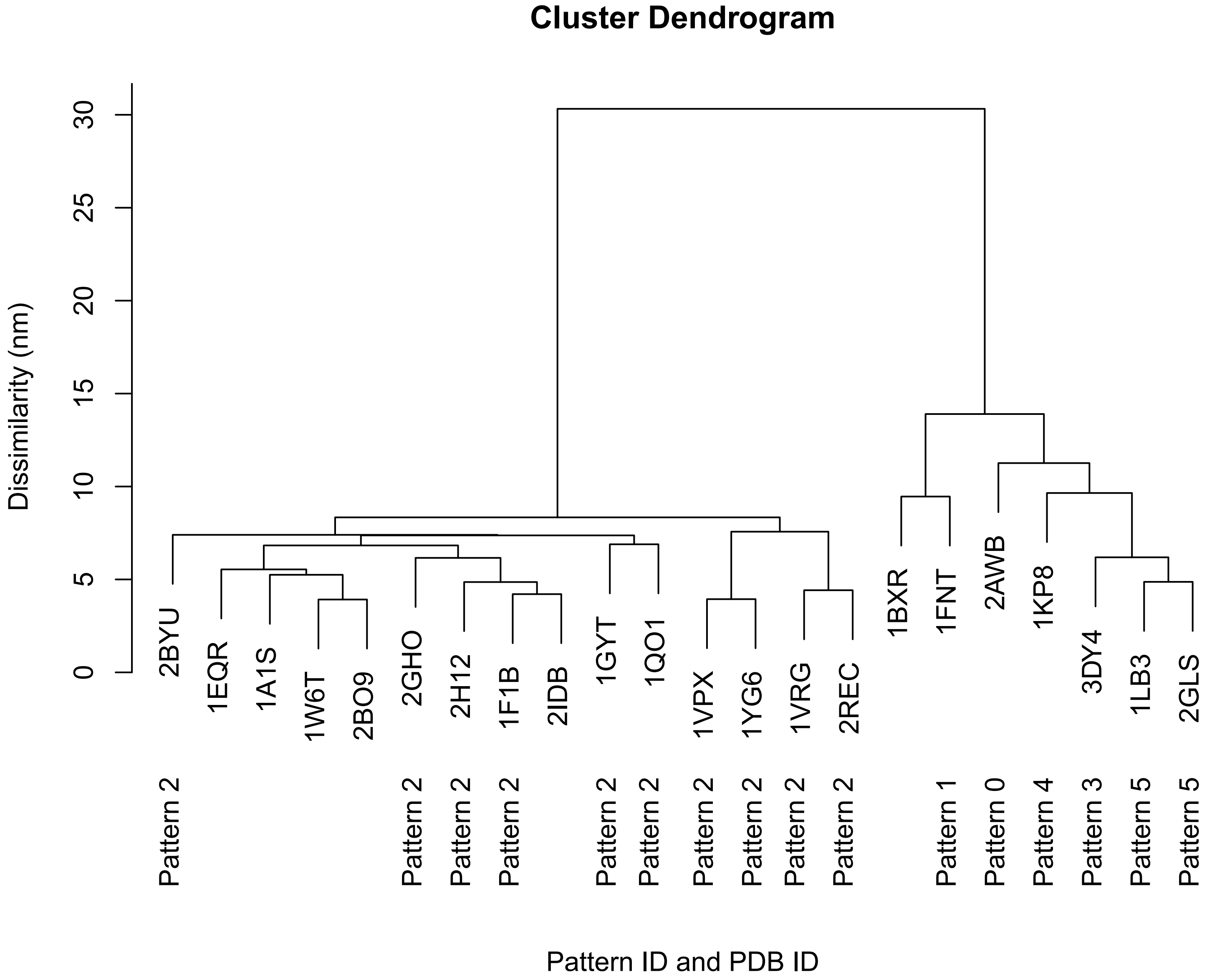

Supplementary Figure 10: Dendrogram of hierarchical clustering of the templates of macromolecular complexes used for simulation, labeled by both PDB ID of the complexes and the ID of patterns whose subtomograms contain these complex. The hierarchical clustering is based on structural dissimilarity in terms of FSC at 0.5 cutoff.

### 2.1.4.4 Additional tests

**2.1.4.4.1 Repeated test** In order to test reproducibility of MPP, we have generated another tomogram with same total number of particles and same copy number frequency for different complexes. The only difference is the localization and orientation of the particles. We then applied same particle picking and MPP procedure to this data. The particle picking step extracted 4,888 subtomograms, among which 4,642 are assigned true class labels.

The result of MPP is summarized in Supplementary Figure 11 and Supplementary Data 5. In terms of predicted patterns, the test results is very similar with the test result in Supplementary Note 2.1.4.3, showing that the recovery of distinct structures through MPP is highly reproducible. MPP in both tests can correctly predict the 1KP8, 2AWB, and 3DY4 in terms of both structural and subtomogram membership consistency. In addition, 1LB3 and 2GLS are correctly predicted. On the other hand, due to the limited ability of redundancy removal when sample number is small, 1FNT are repeatedly predicted in two patterns (ID: 2 and 6). In addition, similar to the test in Supplementary Note 2.1.4.3, MPP also predicted a pattern (ID 1) whose subtomograms are mixed of very similar set of complexes.

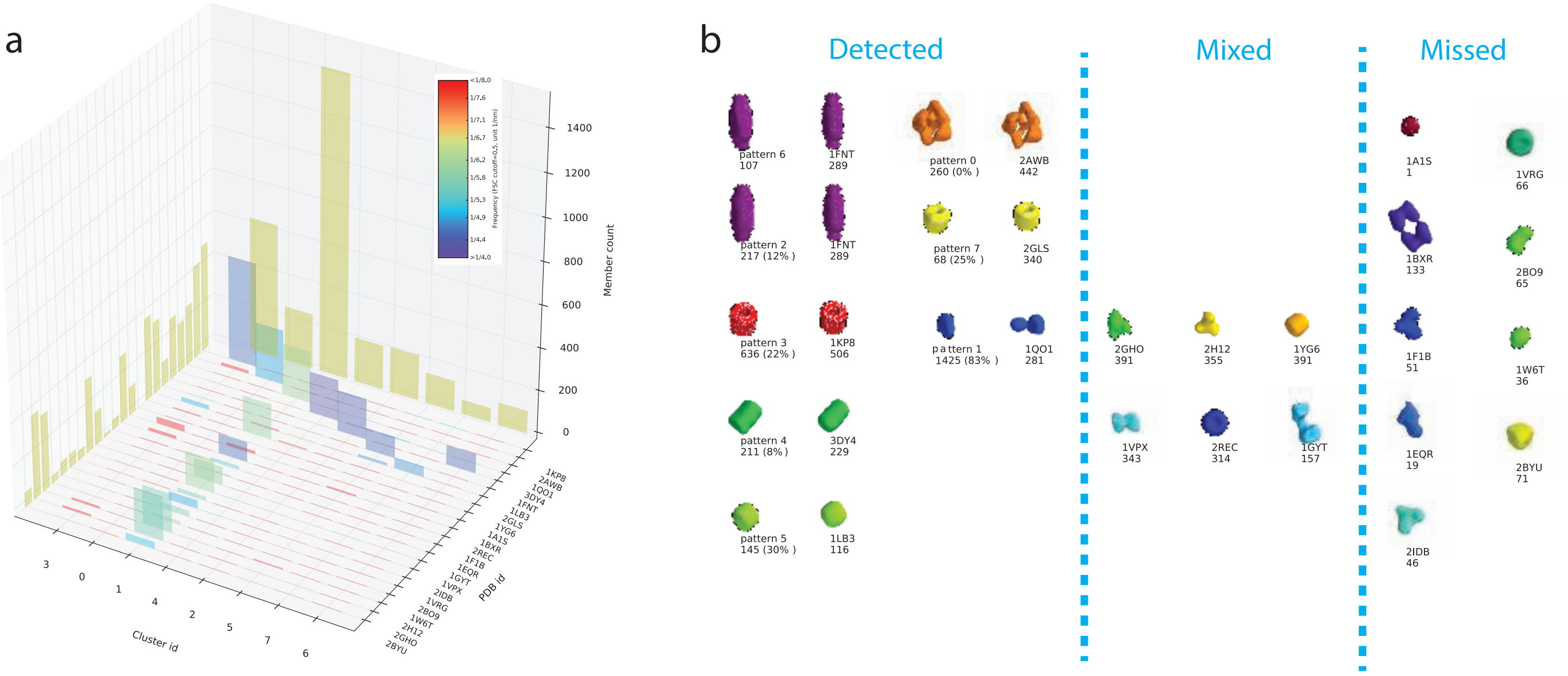

Supplementary Figure 11: Comparison of MPP pattern mining results with ground truth complexes for tomograms containing a crowded mixture of complexes. (a) Column plot representation of the contingency table (Supplementary Data 5) of the subtomogram membership overlap between true and inferred patterns. The height of each column corresponds to the number of subtomograms of the ground truth complex and the predicted patterns. The colors of the columns indicate the structural consistency between the ground truth and the corresponding pattern averages, quantified as FSC with cutoff 0.5 (Supplementary Section2.1.2). (b) The isosurfaces of predicted patterns compared to ground truth structures. The ground truth structures are indicated by their PDB ID code, and the number of instances. Also shown are the isosurface representations of the predicted patterns with the number of instances and the false discovery rate (FDR) in parentheses. The FDR shows the fraction of wrongly assigned subtomograms in the pattern.

**2.1.4.4.2 Test on less crowded tomogram** We also considered the case that tomograms contains less crowded protein complex mixture. Such tomograms may be obtained from cell lysate samples [7, 12, 71, 1, 82, 43, 14, 49]. In this case, we pack 10,000 particles into two volumes, each with size $600 \times 600 \times 200\text{nm}^3$ containing 5,000 particles. We then generate simulated tomograms (Supplementary Note 2.1.1) with tilt angle range ±60° and SNR 50 (Supplementary Figure 12). The crowding level (volume occupancy) is 7.4%. We process the tomograms as described in Supplementary Note 2.1.4.2. The particle picking step extracted 6,802 subtomograms, among which 5,963 are assigned true class labels.

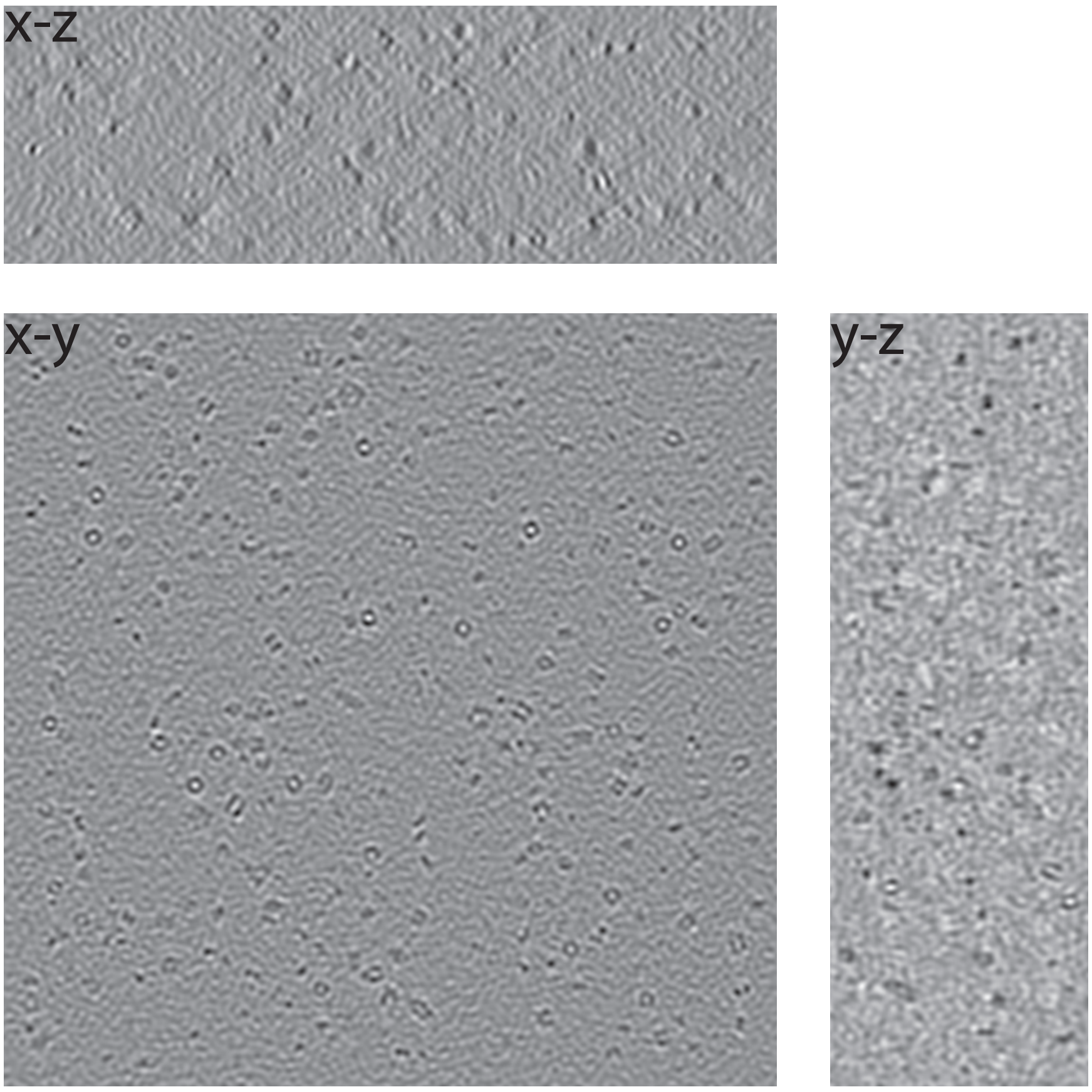

Supplementary Figure 12: Three slices of the simulated tomogram.

We apply MPP on these subtomograms. The result is summarized in Supplementary Figure 13 and Supplementary Data 6. Because the process is affected by less amount of neighboring structures, compared to the result on crowded tomograms (Sections 2.1.4.3 and 2.1.4.4.1), MPP achieved better performance. It has detected more patterns (10 patterns in total). Subtomograms of 18 complexes are generally included inside no more than one patterns, indicating a low redundancy. There are 4 patterns (ID: 0, 1, 5, 9) uniquely correspond to 4 complexes (PDB ID: 2AWB, 1BXR, 1FNT, 1VPX) respectively. There are three patterns (ID: 4, 6, 3), each of which is dominated by subtomograms of one complex (PDB ID: 1KP8, 1LB3, 1YG6), but also mixed with a small portion of another complex with structure consistency $\leq$ 5.0nm. The structure of each of these pattern is most close to the dominant complex with structure consistency $\leq$ 4.5nm. There are 4 complexes (PDB ID: 1YG6, 3DY4, 2GLS, 2BYU) each has subtomograms spread into subtomogram sets of two patterns. Among which, subtomograms of one complex (PDB ID: 3DY4) are distributed into two patterns (ID: 7, 8) that are specific to this complex (FDR $\leq$ 1%). There is also a pattern (ID 2) whose subtomograms contain at least 9 complexes with relately small size, among which subtomograms of at least 7 complexes (PDB ID: 1VRG, 1QO1, 1EQR, 2IDB, 2BO9, 2H12, 2GHO) are mainly inside subtomograms of only this pattern.

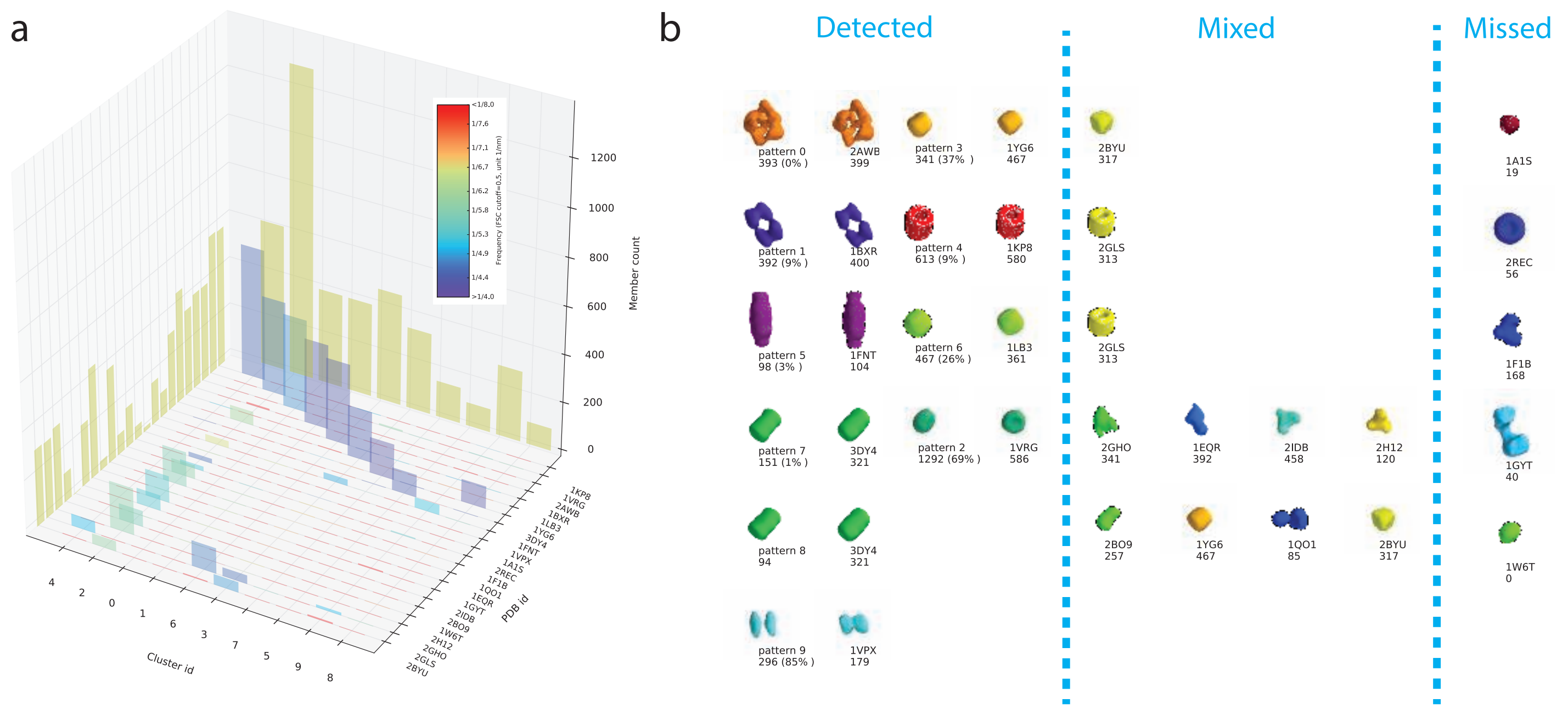

Supplementary Figure 13: Comparison of MPP pattern mining results with ground truth complexes for tomograms containing a crowded mixture of complexes. (a) Column plot representation of the contingency table (Supplementary Data 6) of the subtomogram membership overlap between true and inferred patterns. The height of each column corresponds to the number of subtomograms of the ground truth complex and the predicted patterns. The colors of the columns indicate the structural consistency between the ground truth and the corresponding pattern averages, quantified as FSC with cutoff 0.5 (Supplementary Section2.1.2). (b) The isosurfaces of predicted patterns compared to ground truth structures. The ground truth structures are indicated by their PDB ID code, and the number of instances. Also shown are the isosurface representations of the predicted patterns with the number of instances and the false discovery rate (FDR) in parentheses. The FDR shows the fraction of wrongly assigned subtomograms in the pattern.

## 2.2 Subtomograms extracted from experimental cellular tomograms

### 2.2.1 *A. longum* cell

**2.2.1.1 Tomogram acquisition** *A. longum* cells were frozen and imaged as described previously [90, 89]. Data were collected from $-65°$ to $65°$, with an angular step of $1°$, a total dose of $200e^{-}/\text{Å}^{2}$, a defocus value of -10$\mu$m, and a pixel size of 1.2 nm on a 300-kV FEG G2 Polara transmission electron microscope (TEM) equipped with a lens-coupled 4k-by-4k Ultracam (Gatan, CA) and an energy filter. Data were collected automatically with the UCSF tomography package [108] and reconstructed using the IMOD software package [45] (Supplementary Figure 14).

**2.2.1.2 Particle picking and subtomogram extraction** For particle picking, we filtered the tomogram using the DoG function with $\sigma = 7$nm. We then select the top 10,000 peaks and remove those peaks at the boundary of the tomogram. Centered at each peak we extract a subtomogram of size $18^3$ voxels. This procedure resulted in 9,703 extracted subtomograms.

**2.2.1.3 Pattern mining** We first perform level set based pose normalization (Supplementary Note 1.2) . Then we perform k-means clustering on pose normalized subtomograms to separate the subtomograms into 100 clusters (Supplementary Data 7a). A slight variation of the cluster number does not change the outcome of our analysis. Then based on the shape of the cluster centers, we manually select and combine a number of clusters into 5 groups whose averages are similar (of similar sizes) (Supplementary Data 7a).

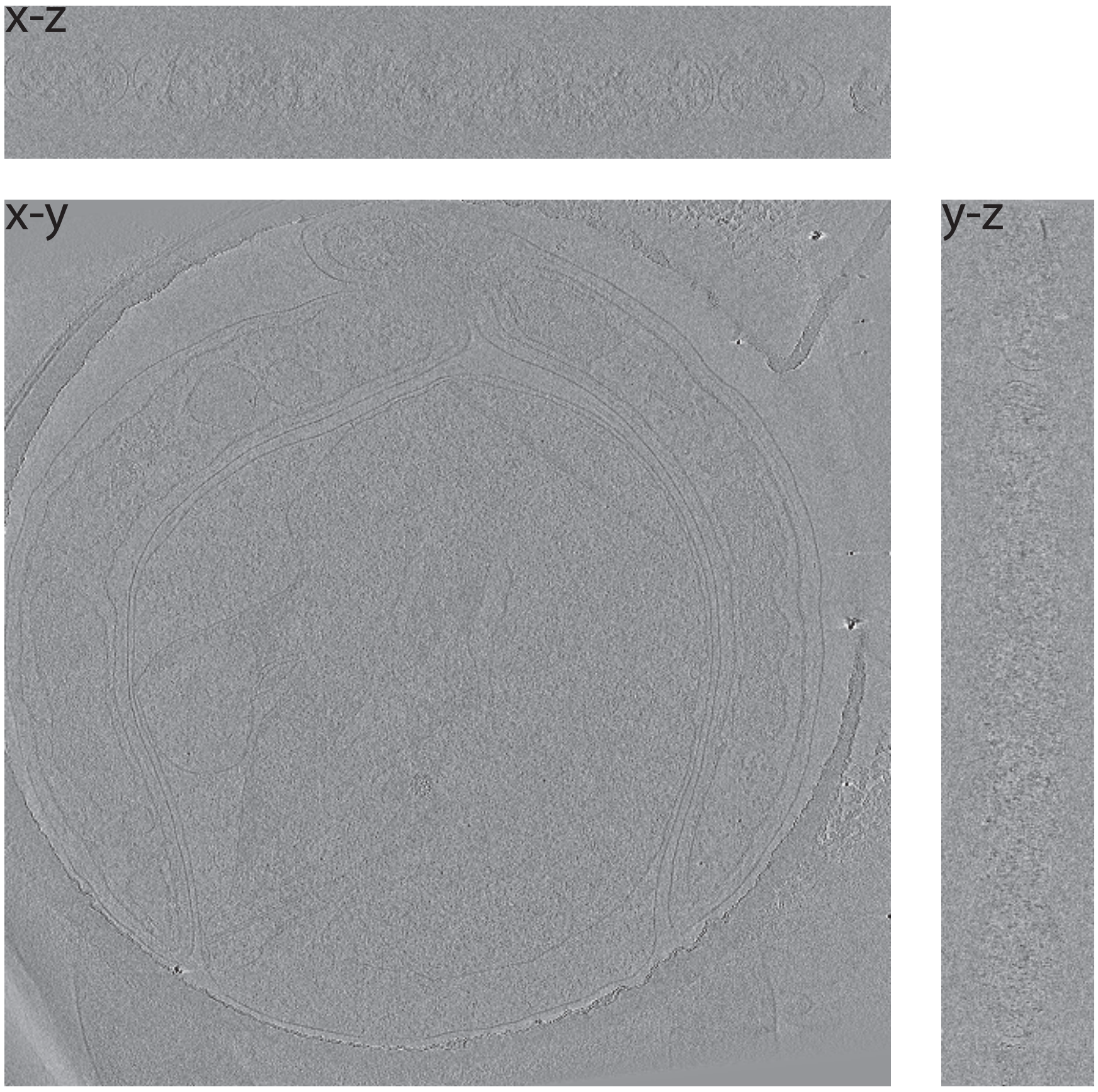

Supplementary Figure 14: Three slices of the tomogram.

We then applied MPP (Supplementary Note 1.1.1) to subtomograms in each group, with random initial orientations, and an initial $k^{\text{k_means_fix}} = 10$ clusters. A slight variation of $k^{\text{k_means_fix}}$ does not change the outcome of our analysis. The resulting predicted patterns are summarized in Supplementary Data 7b and Supplementary Figure 15. Pattern 4 had a structure similar to the GroEL complex (**Figure 3c** in the main text, Supplementary Figure 16). For this pattern, we applied our GA based refinement of subtomogram membership (Supplementary Note 1.1.3.3). The resulting predicted patterns are summarized in Supplementary Data 7b and Supplementary Figure 15.

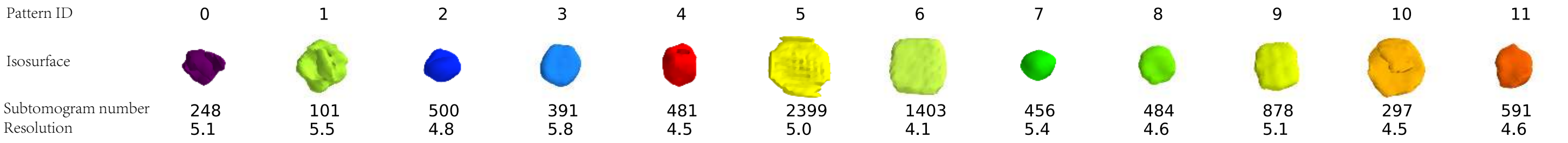

Supplementary Figure 15: ID, isosurface, instance number and resolution of detected patterns. Pattern 4 is likely to be GroEL complex. Patterns 5, 6, and 9 are likely to be membrane patterns.

**2.2.1.4 Validating pattern 4 using template search** Pattern 4 (resolution 4.5nm, Supplementary Data 7b) had a shape and size similar to the GroEL complex. We further assess the likelihood of pattern 4 representing the GroEL complex. We aligned all extracted 9,703 subtomograms against a collection of 28 complexes at 4nm resolution.

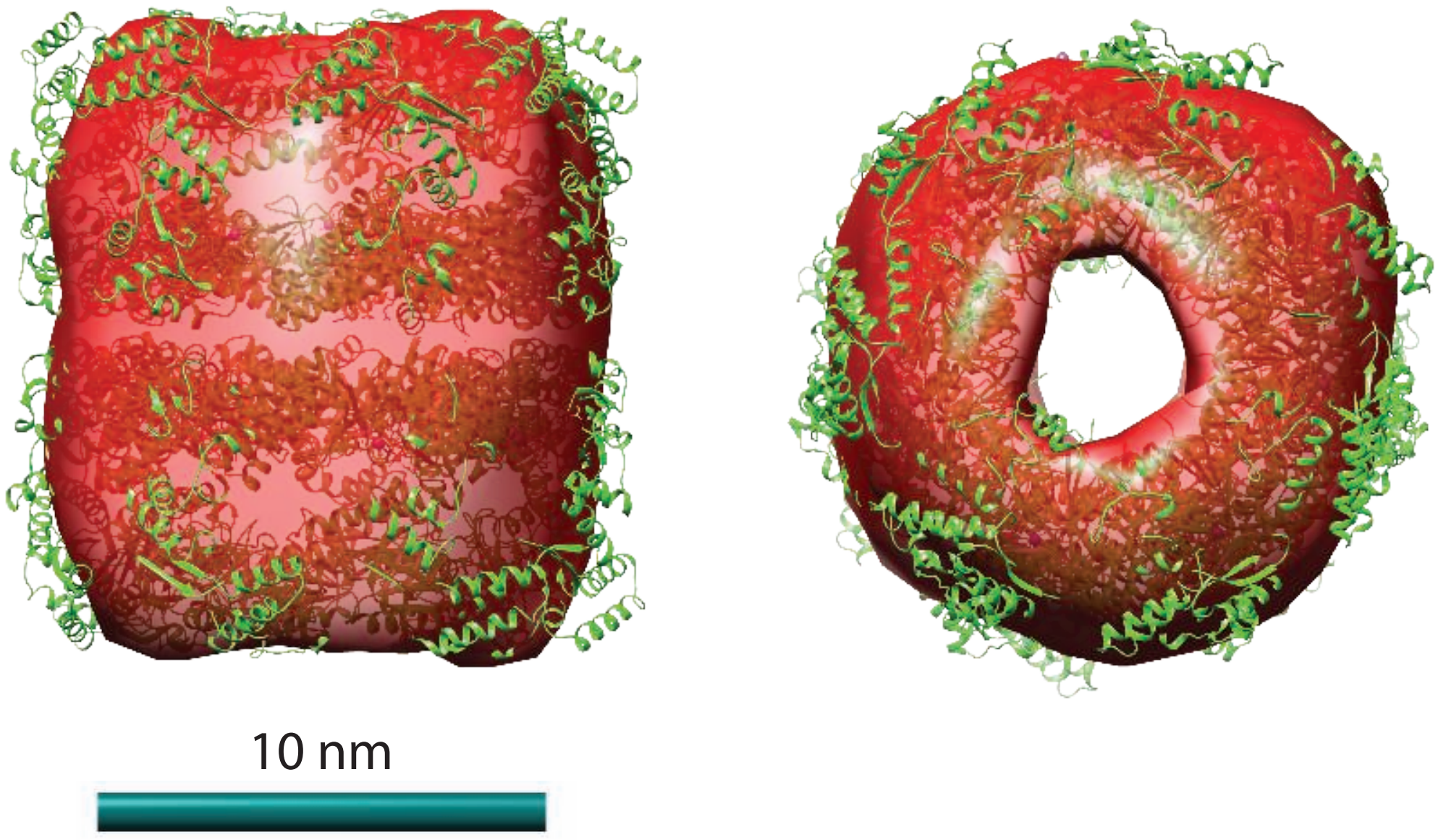

Supplementary Figure 16: The isosurface of the subtomogram average density map of pattern 4, fitted with the known structure of the GroEL (PDB ID: 1KP8).

We aligned all subtomograms in pattern 4 against a collection of possible template structures from the PDB. The subtomograms of pattern 4 were statistically significant most similar to the GroEL template (PDB ID: 1KP8) with respect to the distribution of alignment scores (Supplementary Figure 17a) than to any other template. The second most similar complex is GroEL/GroES template (PDB ID: 1AON). The alignment scores of subtomgrams of pattern 4 are statistically significant higher compared to those of all other extracted subtomograms when aligned to the GroEL template complexes (Supplementary Figure 17b). This observation indicates that there is a high chance for subtomograms of pattern 4 to be among the highest matches and despite being a reference-free method our MPP method therefore performs similarly well to a template search using the GroEL complex as template. Finally, we compare how well subtomograms in the other detected patterns match against the GroEL template. Again, among all predicted patterns the subtomograms of pattern 4 have highest alignment scores against GroEL (PDB ID: 1KP8) (Supplementary Figure 17c). All these observations support the hypothesis that what the subtomograms of pattern 4 contain is most likely a GroEL analog.

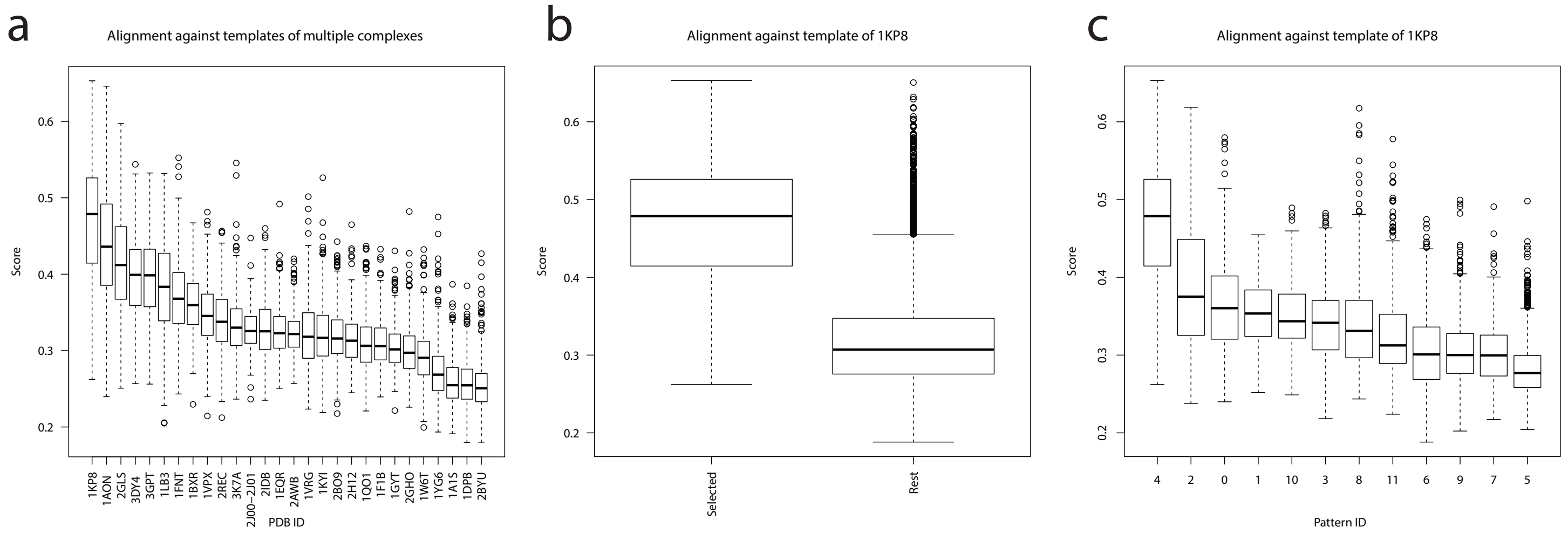

Supplementary Figure 17: (a) Box plot of the distribution of alignment scores of the subtomograms of pattern 4 against all different template complexes (denoted by PDB ID). The complexes are ordered according to median score in descending order. (b) (left) Box plot of the alignment score distribution of subtomograms in pattern 4 against the GroEL template complex (PDB ID: 1KP8) and (right) box plot of the alignment score distribution of all other extracted subtomograms against the GroEL template. (c) Box plot of alignment score distributions of the subtomograms in all patterns against the GroEL template (PDB ID: 1KP8). The patterns are ordered according to median score in descending order.

**2.2.1.5 Estimation of SNR level of subtomograms** We estimated the SNR level of subtomograms of pattern 4 using the same procedure as in Supplementary Note 2.1.4.3.1. Such procedure gives an SNR estimate of $0.24 \pm 0.10$ over all subtomograms pairs.

### 2.2.2 *H. gracilis* cell

*H. gracilis* cells were grown 48 hr in ATCC #233 Broth (ATCC, Manassas, VA) to $OD_{600} = 0.1$. 10 nm colloidal gold (Sigma-Aldrich, St. Louis, MO) pretreated with bovine serum albumin was added to the cells to serve as fiducial markers during tomogram reconstruction. 3 $\mu$l of the resulting sample was pipetted onto a freshly glow-discharged Quantifoil copper R2/2 200 EM grid (Quantifoil Micro Tools GmbH, Jena, Germany) and plunge-frozen in a liquid ethane propane mixture using an FEI Vitrobot mark-III (FEI Company, Hillsboro, OR). The frozen grid was then imaged in an FEI Tecnai G2 Polara 300 keV field emission transmission electron microscope (FEI Company, Hillsboro, OR) equipped with a Gatan energy filter (Gatan, Pleasanton, CA) and a Gatan K2 Summit direct detector (Gatan, Pleasanton, CA) at the California Institute of Technology. Energy-filtered tilt series of images of the cell were collected automatically from 60° to +60° at 1° intervals using the UCSF Tomography data collection software [108] with total dosage of 75 $e^{-}/\text{Å}^2$, a defocus of -15 $\mu$m and a pixel size of 4.9Å. The images were aligned and subsequently reconstructed into a tomogram by weighted back-projection method using the IMOD software package [45] (Supplementary Figure 18).

Particle picking was performed as described in Supplementary Note 2.2.1.2. The interior cell regions are manually segmented using the Amira software (Mercury Computer Systems), and the peaks outside the cell regions are excluded.

The pattern mining and pattern validation follow similar procedures as described in Supplementary Note 2.2.1.3, and Supplementary Note 2.2.1.4. The resulting predicted patterns are summarized in Supplementary Data 8b and Supplementary Figure 19.

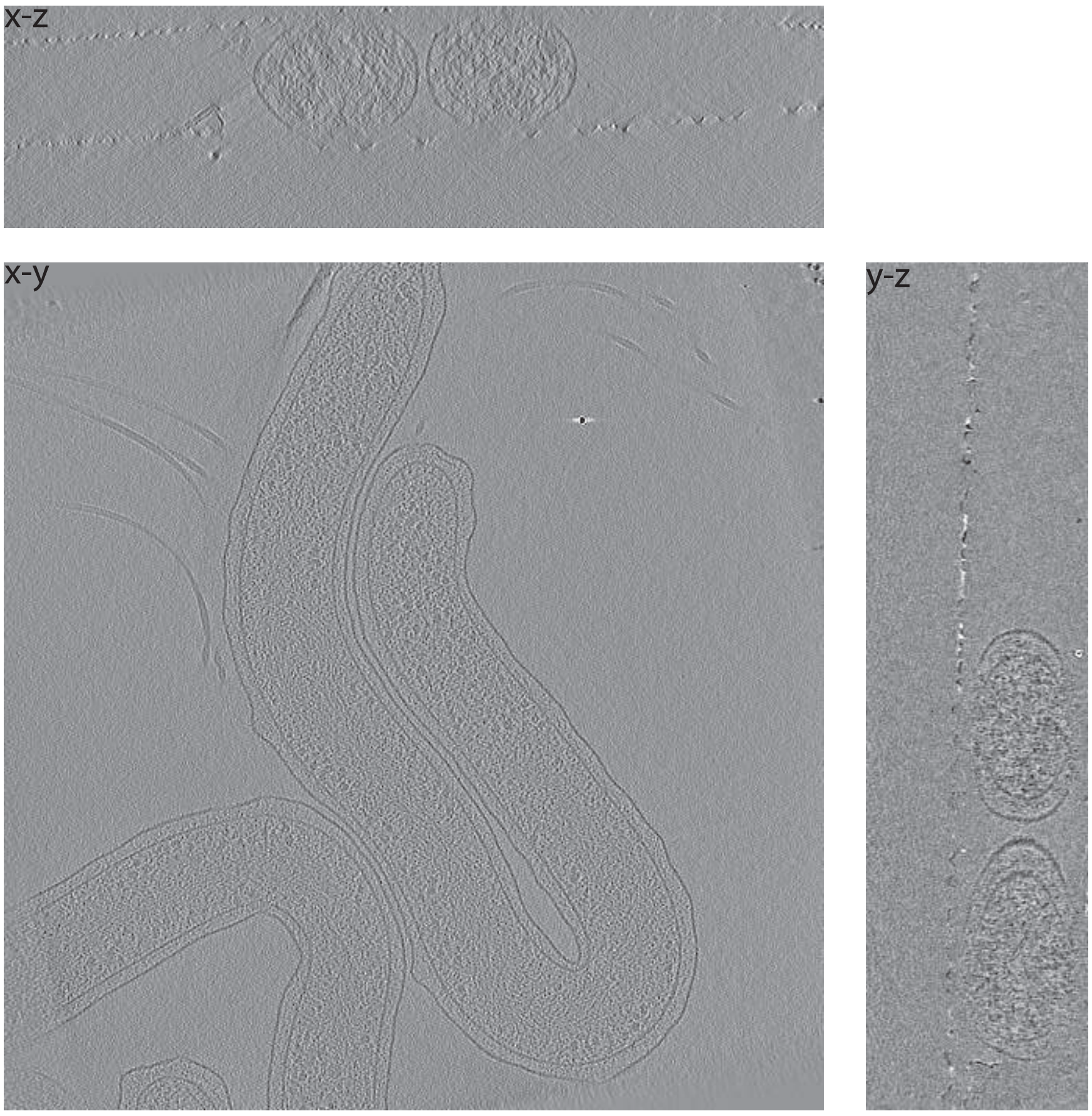

Supplementary Figure 18: Three slices of the tomogram.

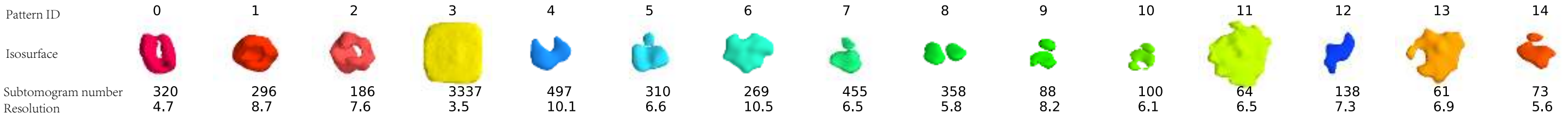

Supplementary Figure 19: ID, isosurface, instance number and resolution of detected patterns. Patterns 0, 1, and 2 are likely to be Ribosome complex. Patterns 3, 11, and 13 are likely to be membrane patterns.

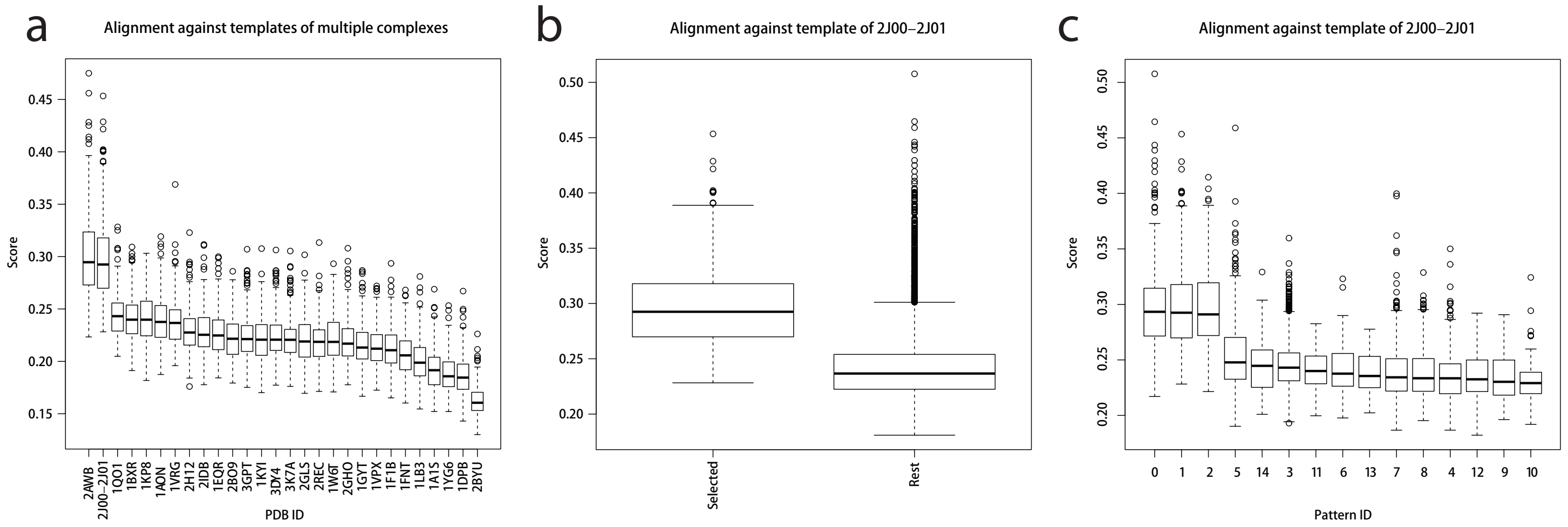

Supplementary Figure 20: (a) Box plot of the distribution of alignment scores of the subtomograms of pattern 1 against all different templates complexes (denoted by PDB ID). The complexes are ordered according to median score in descending order. (b) (left) Box plot of the alignment score distribution of subtomograms in pattern 1 against the ribosome template complex (PDB ID: 2J00-2J01) and (right) box plot of the alignment score distribution of all other extracted subtomograms against the ribosome template, (c) Box plot of alignment score distributions of the subtomograms in all patterns against the ribosome complex template (PDB ID: 22J00-2J01). The patterns are ordered according to median score in descending order.

### 2.2.3 *B. bacteriovorus* cell

*B. bacteriovorus* HD100 cells were grown as described previously [48] on *E. coli* S17-1 prey cells in Ca-HEPES buffer at 29°C until most prey cells were cleared from the culture. 10 nm colloidal gold (Sigma-Aldrich, St. Louis, MO) pretreated with bovine serum albumin was added to the cells to serve as fiducial markers during tomogram reconstruction. 3 $\mu$l of the resulting sample was pipetted onto a freshly glow-discharged Quantifoil copper R2/2 200 EM grid (Quantifoil Micro Tools GmbH, Jena, Germany) and plunge-frozen in a liquid ethane propane mixture using an FEI Vitrobot mark-III (FEI Company, Hillsboro, OR). The frozen grid was then imaged in an FEI Titan Krios 300 keV field emission transmission electron microscope (FEI Company, Hillsboro, OR) equipped with a Gatan energy filter (Gatan, Pleasanton, CA) and a Gatan K2 Summit direct detector (Gatan, Pleasanton, CA) at the Howard Hughes Medical Institute Janelia Research Campus. Energy-filtered tilt series of images of the cell were collected automatically from 65° to +65° at 1° intervals using the UCSF Tomography data collection software[108] with total dosage of 100 $e^-/\text{Å}^2$, a defocus of -8 $\mu$m and a pixel size of 4.2 Å. The images were aligned and subsequently reconstructed into a tomogram by weighted back-projection method using the IMOD software package[45].

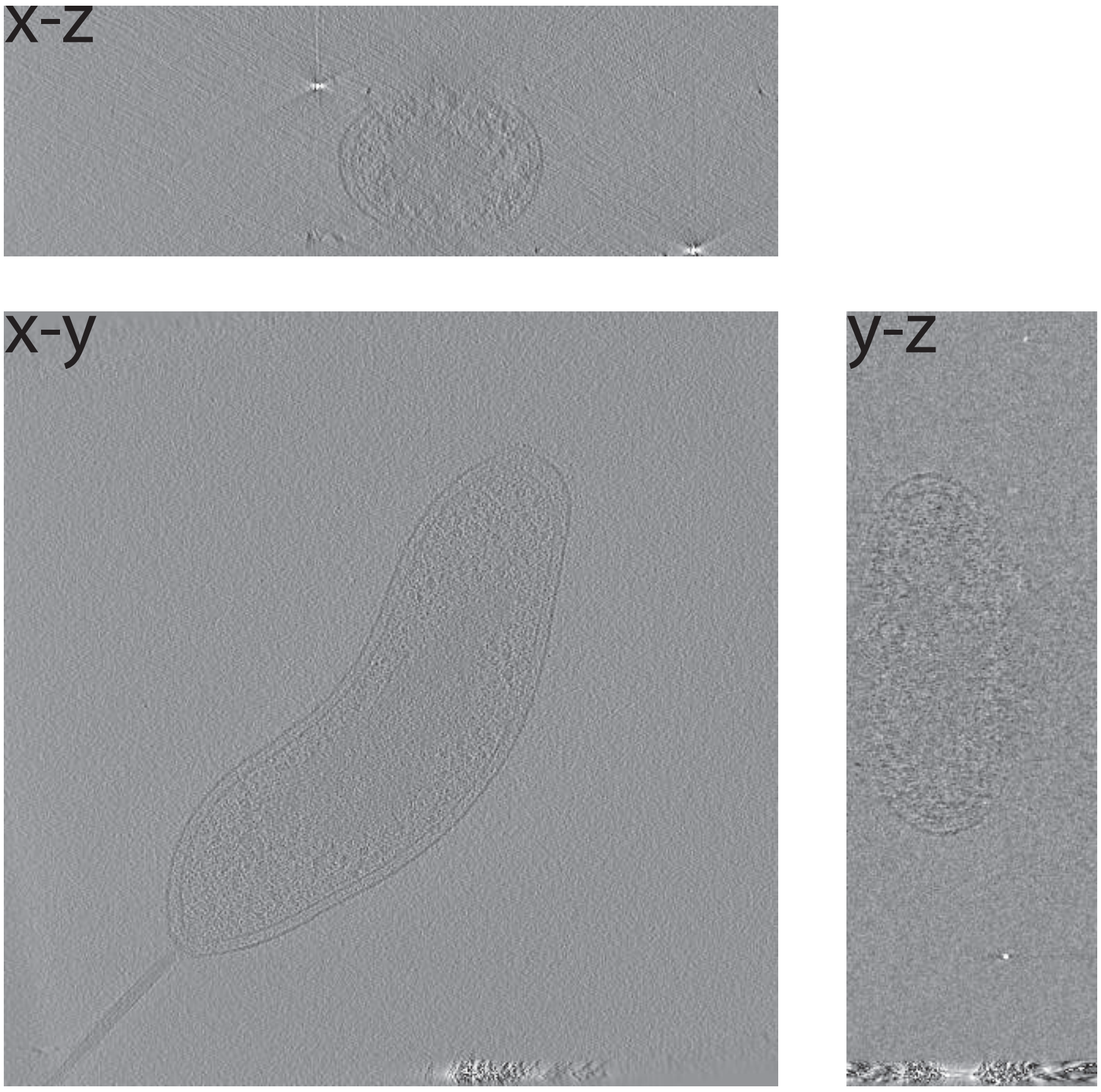

Supplementary Figure 21: Three slices of the tomogram.

Particle picking was performed similar as described in Supplementary Note 2.2.1.2. The interior cell regions were manually segmented using the Amira software (Mercury Computer Systems), and the peaks outside the cell regions are excluded.

The pattern mining and pattern validation were performed similar as described in Supplementary Note 2.2.1.3, and Supplementary Note 2.2.1.4. The resulting predicted patterns are summarized in Supplementary Data 9b and Supplementary Figure 22.

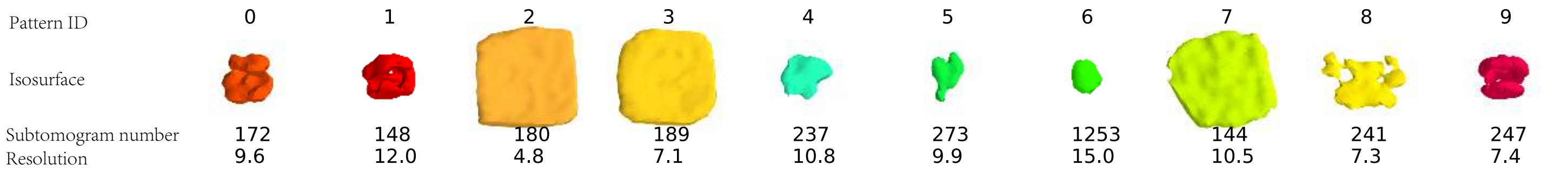

Supplementary Figure 22: ID, isosurface, instance number and resolution of detected patterns. Patterns 0, 1, and 9 are likely to be Ribosome complex. Patterns 2, 3, and 7 are likely to be membrane patterns.

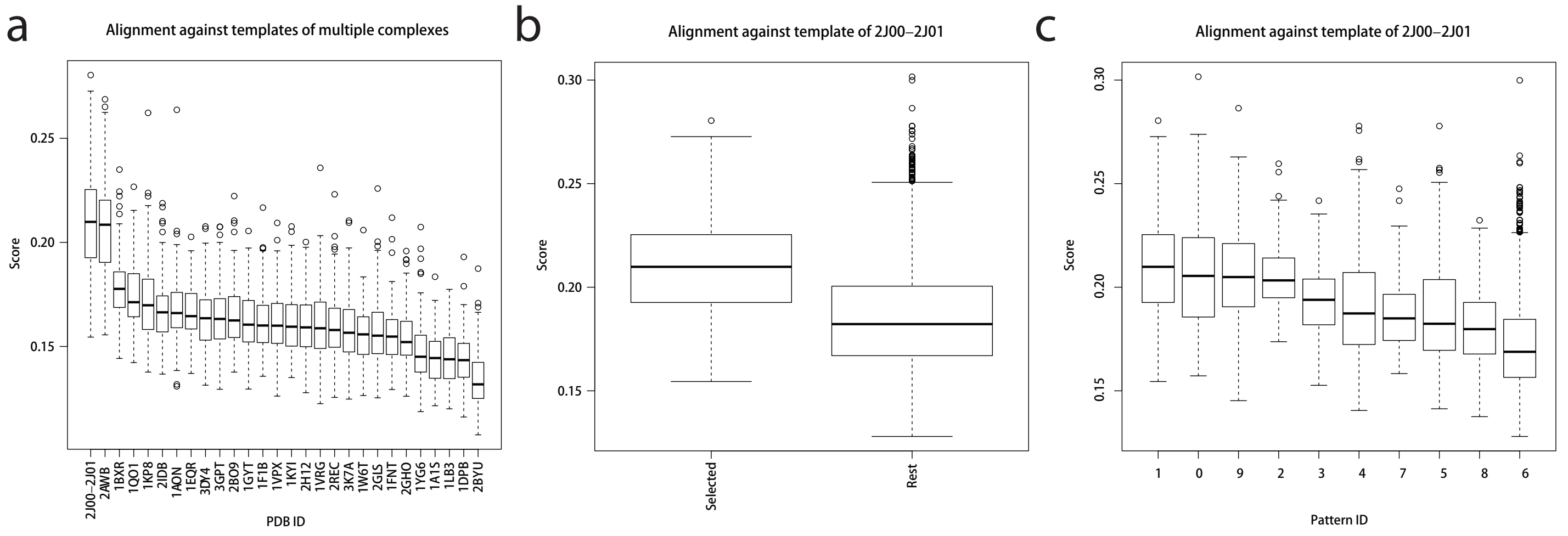

Supplementary Figure 23: (a) Box plot of the distribution of alignment scores of the subtomograms of pattern 1 against all different templates complexes (denoted by PDB ID). The complexes are ordered according to median score in descending order. (b) (left) Box plot of the alignment score distribution of subtomograms in pattern 1 against the ribosome template complex (PDB ID: 2J00-2J01) and (right) box plot of the alignment score distribution of all other extracted subtomograms against the ribosome template, (c) Box plot of alignment score distributions of the subtomograms in all patterns against the ribosome complex template (PDB ID: 22J00-2J01). The patterns are ordered according to median score in descending order.

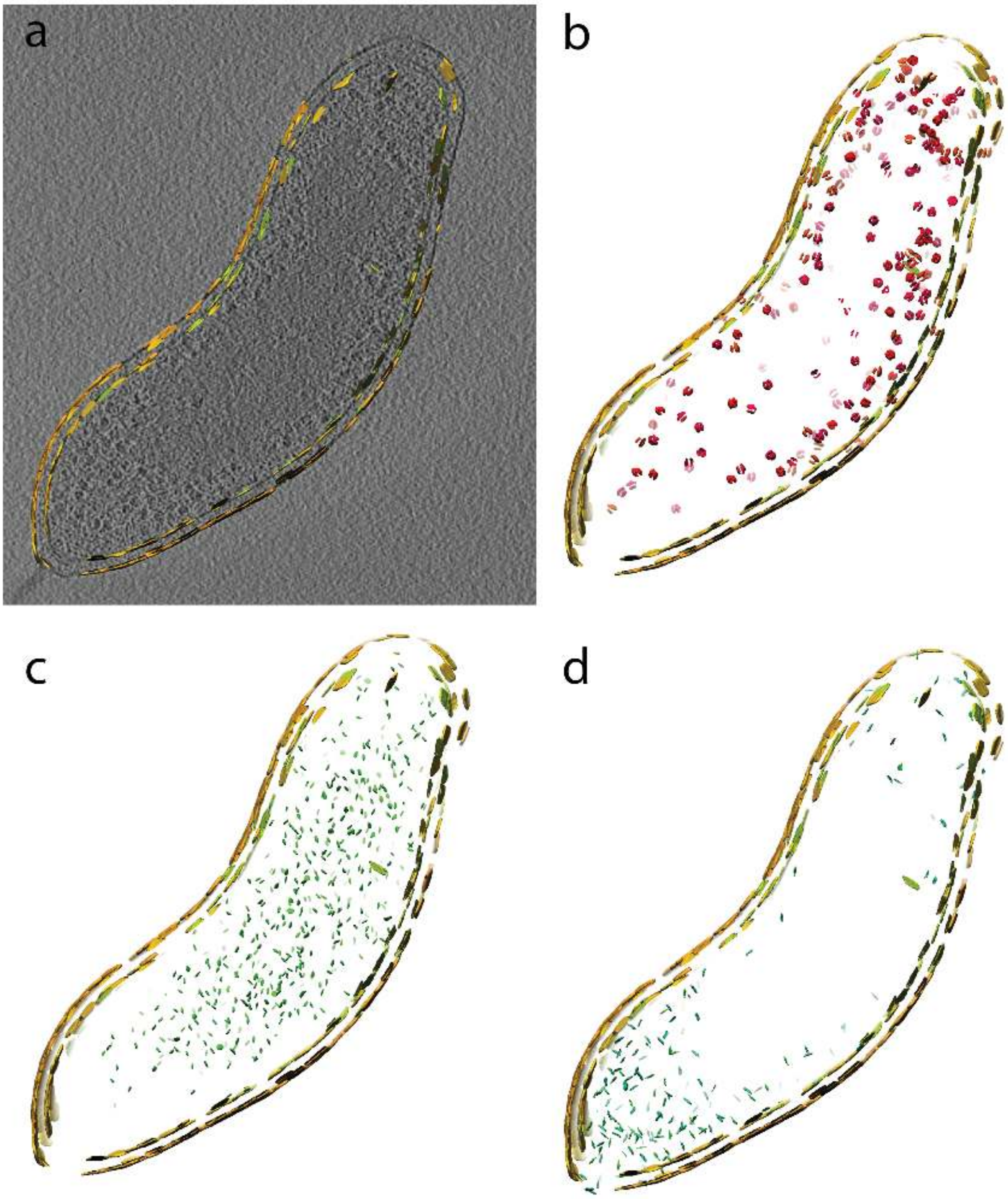

Supplementary Figure 24: A thin section of embedded instances of different patterns, outlined by embedded instances of membrane patterns. (a) A slice of tomogram. (b) Patterns 0, 1, and 9. (c) Pattern 6. (d) Patterns 4 and 5.